\RequirePackage{fix-cm}
\documentclass[nospthms,epjc3,final]{svjour3}  
\smartqed  % flush right qed marks, e.g. at end of proof
\RequirePackage{graphicx}
\journalname{Eur. Phys. J. C}
\usepackage{mathtools,amsmath,amssymb,amsfonts,dsfont,mathrsfs,amsthm,latexsym}

\usepackage{bm}
\usepackage{graphicx}
\usepackage{centernot}
\usepackage{xcolor}
\usepackage{comment}
\usepackage{feynmf}
\usepackage{siunitx}
\usepackage{array}
\newcolumntype{L}[1]{>{\raggedright\let\newline\\\arraybackslash\hspace{0pt}}m{#1}}
\newcolumntype{C}[1]{>{\centering\let\newline\\\arraybackslash\hspace{0pt}}m{#1}}
\newcolumntype{R}[1]{>{\raggedleft\let\newline\\\arraybackslash\hspace{0pt}}m{#1}}
\usepackage{tikz}
\usepackage{pgfplots}
\pgfplotsset{width=7cm}
\usepackage{braket}
\usepackage{xparse}
\usepackage{breqn}
\usetikzlibrary{
  arrows.meta,
  shapes,
  cd,
  spy,
  decorations,
  decorations.pathmorphing,
  decorations.markings,
  calc,
  shadows.blur,
  shadings,
  trees}

\tikzset{
  Gr/.style={circle,draw,minimum size=8mm},
  Ga/.style={circle,draw,minimum size=8mm,fill=black!10,delta angle=40},
}

\newcommand{\A}{\mathcal{A}} %% This is equivalent to \Ag (no args)

\DeclareDocumentCommand{\Ag}{ s o }{ \IfBooleanTF{#1}
    { \IfValueTF{#2}{ \bm{\mathcal{A}}_{(#2)} }{ \bm{\mathcal{A}} } }
    { \IfValueTF{#2}{    {\mathcal{A}}_{(#2)} }{    {\mathcal{A}} } } }
\DeclareDocumentCommand{\Af}{ s o }{ \IfBooleanTF{#1}
    { \IfValueTF{#2}{ \boldsymbol{A}_{(#2)} }{ \boldsymbol{A} } }
    { \IfValueTF{#2}{            {A}_{(#2)} }{            {A} } } }
\DeclareDocumentCommand{\Fg}{ s o }{ \IfBooleanTF{#1}
    { \IfValueTF{#2}{ \bm{\mathcal{F}}_{(#2)} }{ \bm{\mathcal{F}} } }
    { \IfValueTF{#2}{    {\mathcal{F}}_{(#2)} }{    {\mathcal{F}} } } }
\DeclareDocumentCommand{\Ff}{ s o }{ \IfBooleanTF{#1}
    { \IfValueTF{#2}{ \boldsymbol{F}_{(#2)} }{ \boldsymbol{F} } }
    { \IfValueTF{#2}{            {F}_{(#2)} }{            {F} } } }
\newcommand{\Hi}{\mathcal{H}}

\newcommand{\Lag}{\mathscr{L}}

\newcommand{\Mi}{\mathcal{M}}

\DeclareDocumentCommand{\PB}{ O{m} O{q} O{p} m m }{ \frac{ \partial #4 }{\partial {#2}^{#1} } \frac{ \partial #5 }{\partial {#3}_{#1} } - \frac{ \partial #4 }{\partial {#3}_{#1} } \frac{ \partial #5 }{\partial {#2}^{#1} } }

\newcommand{\R}{\mathbb{R}}
\newcommand{\Ri}{\mathcal{R}}
\DeclareDocumentCommand\Te{o o O{\,} m }{{T}_{#3}{}^{#1}_{#2}(#4)}

\DeclareDocumentCommand{\BM}{ s }{ \IfBooleanTF{#1} {\hat{\bm{M}}}{\bm{M}} }
\DeclareDocumentCommand{\BN}{ s }{ \IfBooleanTF{#1} {\hat{\bm{N}}}{\bm{N}} }
\DeclareDocumentCommand{\BP}{ s }{ \IfBooleanTF{#1} {\hat{\bm{P}}}{\bm{P}} }
\DeclareDocumentCommand{\BQ}{ s }{ \IfBooleanTF{#1} {\hat{\bm{Q}}}{\bm{Q}} }
\DeclareDocumentCommand{\BR}{ s }{ \IfBooleanTF{#1} {\hat{\bm{R}}}{\bm{R}} }
\DeclareDocumentCommand{\BS}{ s }{ \IfBooleanTF{#1} {\hat{\bm{S}}}{\bm{S}} }
\DeclareDocumentCommand{\BU}{ s }{ \IfBooleanTF{#1} {\hat{\bm{U}}}{\bm{U}} }
\DeclareDocumentCommand{\BV}{ s }{ \IfBooleanTF{#1} {\hat{\bm{V}}}{\bm{V}} }

\NewDocumentCommand\MyAc{ m }{#1}
\DeclareDocumentCommand{\vif}{ t. t, t- s s m }{
  \RenewDocumentCommand\MyAc{ m }{##1}
  \IfBooleanT{#1}{\RenewDocumentCommand\MyAc{ m }{ \mathring{##1} } }
  \IfBooleanT{#2}{\RenewDocumentCommand\MyAc{ m }{ \tilde{##1} } }
  \IfBooleanT{#3}{\RenewDocumentCommand\MyAc{ m }{ \bar{##1} } }
  \IfBooleanTF{#4}
  { \IfBooleanTF{#5} { \hat{\MyAc{\boldsymbol{e}}}^{\hat{#6}} }{ \hat{\MyAc{\boldsymbol{e}}}^{{#6}} } }
  { \MyAc{\boldsymbol{e}}^{{#6}} } }
\DeclareDocumentCommand{\vi}{ t. t, t- s s m m}{
  \RenewDocumentCommand\MyAc{ m }{##1}
  \IfBooleanT{#1}{\RenewDocumentCommand\MyAc{ m }{ \mathring{##1} } }
  \IfBooleanT{#2}{\RenewDocumentCommand\MyAc{ m }{ \tilde{##1} } }
  \IfBooleanT{#3}{\RenewDocumentCommand\MyAc{ m }{ \bar{##1} } }
  \IfBooleanTF{#4}
  { \IfBooleanTF{#5} { \hat{\MyAc{e}}^{\hat{#6}}_{\hat{#7}} }{ \hat{\MyAc{e}}^{#6}_{{#7}} } }
  { \MyAc{e}^{{#6}}_{{#7}} } }
\DeclareDocumentCommand{\bt}{ t. t, t- s s m m m }{
  \RenewDocumentCommand\MyAc{ m }{##1}
  \IfBooleanT{#1}{\RenewDocumentCommand\MyAc{ m }{ \mathring{##1} } }
  \IfBooleanT{#2}{\RenewDocumentCommand\MyAc{ m }{ \tilde{##1} } }
  \IfBooleanT{#3}{\RenewDocumentCommand\MyAc{ m }{ \bar{##1} } }
  \IfBooleanTF{#4}
  { \IfBooleanTF{#5} { \hat{\MyAc{\mathcal{B}}}_{{#6}}{}^{\hat{#7}}{}_{\hat{#8}} }{ \hat{\MyAc{\mathcal{B}}}_{{#6}}{}^{{#7}}{}_{{#8}} } }
  { \MyAc{\mathcal{B}}_{{#6}}{}^{{#7}}{}_{{#8}} } }

\DeclareDocumentCommand{\ct}{ t. t, t- s s m m m }{
  \RenewDocumentCommand\MyAc{ m }{##1}
  \IfBooleanT{#1}{\RenewDocumentCommand\MyAc{ m }{ \mathring{##1} } }
  \IfBooleanT{#2}{\RenewDocumentCommand\MyAc{ m }{ \tilde{##1} } }
  \IfBooleanT{#3}{\RenewDocumentCommand\MyAc{ m }{ \bar{##1} } }
  \IfBooleanTF{#4}
  { \IfBooleanTF{#5} { \hat{\MyAc{\Gamma}}_{{#6}}{}^{\hat{#7}}{}_{\hat{#8}} }{ \hat{\MyAc{\Gamma}}_{{#6}}{}^{{#7}}{}_{{#8}} } }
  { \MyAc{\Gamma}_{{#6}}{}^{{#7}}{}_{{#8}} } }
\DeclareDocumentCommand{\spif}{ t. t, t- s s m m }{
  \RenewDocumentCommand\MyAc{ m }{##1}
  \IfBooleanT{#1}{\RenewDocumentCommand\MyAc{ m }{ \mathring{##1} } }
  \IfBooleanT{#2}{\RenewDocumentCommand\MyAc{ m }{ \tilde{##1} } }
  \IfBooleanT{#3}{\RenewDocumentCommand\MyAc{ m }{ \bar{##1} } }
  \IfBooleanTF{#4}
  { \IfBooleanTF{#5} { \hat{\MyAc{\boldsymbol{\omega}}}^{\hat{#6}}{}_{\hat{#7}} }{ \hat{\MyAc{\boldsymbol{\omega}}}^{{#6}}{}_{{#7}} } }
  { \MyAc{\boldsymbol{\omega}}^{{#6}}{}_{{#7}} } }
\DeclareDocumentCommand{\spi}{ t. t, t- s s m m m }{
  \RenewDocumentCommand\MyAc{ m }{##1}
  \IfBooleanT{#1}{\RenewDocumentCommand\MyAc{ m }{ \mathring{##1} } }
  \IfBooleanT{#2}{\RenewDocumentCommand\MyAc{ m }{ \tilde{##1} } }
  \IfBooleanT{#3}{\RenewDocumentCommand\MyAc{ m }{ \bar{##1} } }
  \IfBooleanTF{#4}
  { \IfBooleanTF{#5} { \hat{\MyAc{{\omega}}}_{\hat{#6}}{}^{\hat{#7}}{}_{\hat{#8}} }{ \hat{\MyAc{{\omega}}}_{{#6}}{}^{{#7}}{}_{{#8}} } }
  { \MyAc{{\omega}}_{{#6}}{}^{{#7}}{}_{{#8}} } }
\DeclareDocumentCommand{\rif}{ t. t, t- s s m m }{
  \RenewDocumentCommand\MyAc{ m }{##1}
  \IfBooleanT{#1}{\RenewDocumentCommand\MyAc{ m }{ \mathring{##1} } }
  \IfBooleanT{#2}{\RenewDocumentCommand\MyAc{ m }{ \tilde{##1} } }
  \IfBooleanT{#3}{\RenewDocumentCommand\MyAc{ m }{ \bar{##1} } }
  \IfBooleanTF{#4}
  { \IfBooleanTF{#5} { \hat{\MyAc{\bm{\mathcal{R}}}}{}^{\hat{#6}}{}_{\hat{#7}} }{ \hat{\MyAc{\bm{\mathcal{R}}}}{}^{{#6}}{}_{{#7}} } }
  { \MyAc{\bm{\mathcal{R}}}{}^{{#6}}{}_{{#7}} } }
\DeclareDocumentCommand{\ri}{ t. t, t- s s m m m }{
  \RenewDocumentCommand\MyAc{ m }{##1}
  \IfBooleanT{#1}{\RenewDocumentCommand\MyAc{ m }{ \mathring{##1} } }
  \IfBooleanT{#2}{\RenewDocumentCommand\MyAc{ m }{ \tilde{##1} } }
  \IfBooleanT{#3}{\RenewDocumentCommand\MyAc{ m }{ \bar{##1} } }
  \IfBooleanTF{#4}
  { \IfBooleanTF{#5} { \hat{\MyAc{\mathcal{R}}}_{{#6}}{}^{\hat{#7}}{}_{\hat{#8}} }{ \hat{\MyAc{\mathcal{R}}}_{{#6}}{}^{{#7}}{}_{{#8}} } }
  { \MyAc{\mathcal{R}}_{{#6}}{}^{{#7}}{}_{{#8}} } }
\DeclareDocumentCommand{\kf}{ t. t, t- s s m m }{
  \RenewDocumentCommand\MyAc{ m }{##1}
  \IfBooleanT{#1}{\RenewDocumentCommand\MyAc{ m }{ \mathring{##1} } }
  \IfBooleanT{#2}{\RenewDocumentCommand\MyAc{ m }{ \tilde{##1} } }
  \IfBooleanT{#3}{\RenewDocumentCommand\MyAc{ m }{ \bar{##1} } }
  \IfBooleanTF{#4}
  { \IfBooleanTF{#5} { \hat{\MyAc{\bm{\mathcal{K}}}}^{\hat{#6}}{}_{\hat{#7}} }{ \hat{\MyAc{\bm{\mathcal{K}}}}^{{#6}}{}_{{#7}} } }
  { \MyAc{\bm{\mathcal{K}}}^{{#6}}{}_{{#7}} } }
\DeclareDocumentCommand{\ko}{ t. t, t- s s m m m }{
  \RenewDocumentCommand\MyAc{ m }{##1}
  \IfBooleanT{#1}{\RenewDocumentCommand\MyAc{ m }{ \mathring{##1} } }
  \IfBooleanT{#2}{\RenewDocumentCommand\MyAc{ m }{ \tilde{##1} } }
  \IfBooleanT{#3}{\RenewDocumentCommand\MyAc{ m }{ \bar{##1} } }
  \IfBooleanTF{#4}
  { \IfBooleanTF{#5} { \hat{\MyAc{\mathcal{K}}}_{\hat{#6}}{}^{\hat{#7}}{}_{\hat{#8}} }{ \hat{\MyAc{\mathcal{K}}}_{{#6}}{}^{{#7}}{}_{{#8}} } }
  { \MyAc{\mathcal{K}}_{{#6}}{}^{{#7}}{}_{{#8}} } }
\DeclareDocumentCommand{\tf}{ t. t, t- s s m }{
  \RenewDocumentCommand\MyAc{ m }{##1}
  \IfBooleanT{#1}{\RenewDocumentCommand\MyAc{ m }{ \mathring{##1} } }
  \IfBooleanT{#2}{\RenewDocumentCommand\MyAc{ m }{ \tilde{##1} } }
  \IfBooleanT{#3}{\RenewDocumentCommand\MyAc{ m }{ \bar{##1} } }
  \IfBooleanTF{#4}
  { \IfBooleanTF{#5} { \hat{\MyAc{\bm{\mathcal{T}}}}^{\hat{#6}} }{ \hat{\MyAc{\bm{\mathcal{T}}}}^{{#6}} } }
  { \MyAc{\bm{\mathcal{T}}}^{{#6}} } }
\DeclareDocumentCommand{\tt}{ t. t, t- s s m m m }{
  \RenewDocumentCommand\MyAc{ m }{##1}
  \IfBooleanT{#1}{\RenewDocumentCommand\MyAc{ m }{ \mathring{##1} } }
  \IfBooleanT{#2}{\RenewDocumentCommand\MyAc{ m }{ \tilde{##1} } }
  \IfBooleanT{#3}{\RenewDocumentCommand\MyAc{ m }{ \bar{##1} } }
  \IfBooleanTF{#4}
  { \IfBooleanTF{#5} { \hat{\MyAc{\mathcal{T}}}_{\hat{#6}}{}^{\hat{#7}}{}_{\hat{#8}} }{ \hat{\MyAc{\mathcal{T}}}_{{#6}}{}^{{#7}}{}_{{#8}} } }
  { \MyAc{\mathcal{T}}_{{#6}}{}^{{#7}}{}_{{#8}} } }
\DeclareDocumentCommand{\stt}{ t. t, t- s s m m m }{
  \RenewDocumentCommand\MyAc{ m }{##1}
  \IfBooleanT{#1}{\RenewDocumentCommand\MyAc{ m }{ \mathring{##1} } }
  \IfBooleanT{#2}{\RenewDocumentCommand\MyAc{ m }{ \tilde{##1} } }
  \IfBooleanT{#3}{\RenewDocumentCommand\MyAc{ m }{ \bar{##1} } }
  \IfBooleanTF{#4}
  { \IfBooleanTF{#5} { \hat{\MyAc{\mathcal{S}}}_{\hat{#6}}{}^{\hat{#7}}{}_{\hat{#8}} }{ \hat{\MyAc{\mathcal{S}}}_{{#6}}{}^{{#7}}{}_{{#8}} } }
  { \MyAc{\mathcal{S}}_{{#6}}{}^{{#7}}{}_{{#8}} } }
\newcommand{\tor}{\mathcal{T}}

\NewDocumentCommand{\MyLe}{}{}
\DeclareDocumentCommand{\PG}{ s s O{\Pi} m m m }{
  \RenewDocumentCommand\MyLe{}{\Gamma}
  \IfBooleanT{#1}{\RenewDocumentCommand{\MyLe}{}{ \bt{ }{ }{ } }}
  \IfBooleanT{#2}{\RenewDocumentCommand{\MyLe}{}{\Ag}}
  { {#3}_{\MyLe}{}^{#4}{}_{#5}{}^{#6} }
}

\newcommand{\comm}[2]{\left[#1,#2\right]}

\renewcommand{\set}[1]{\ensuremath{\Set{ #1 }}}

\newcommand{\sgn}{\operatorname{sgn}}

\NewDocumentCommand{\tak}{ s m m}{
  \IfBooleanTF{#1}{ \big( {#2} \big) \big[ {#3} \big] }
              { \big( {#2} \big] \big[ {#3} \big) }
}

\newcommand*{\de}[1]{\mathop{\mathrm{d}#1}\nolimits}% differential, facultative argoment between square parentheses
\newcommand\UTFSM{Departamento de F\'isica, Universidad T\'{e}cnica Federico Santa Mar\'\i a, Casilla 110-V, Valpara\'iso, Chile}
\newcommand\UTFSMmat{Departamento de Matem\'aticas, Universidad T\'{e}cnica Federico Santa Mar\'\i a, Casilla 110-V, Valpara\'iso, Chile}

\newcommand{\UdelaR}{Instituto de F\'isica, Facultad de Ciencias, Igu\'a 4225, esq. Mataojo, 11400 Montevideo, Uruguay.}
\usepackage{cite}
\usepackage[colorlinks=true]{hyperref}
\begin{document}

\title{Aspects of the polynomial affine model of gravity in three dimensions}
\subtitle{With focus in the cosmological solutions}

\titlerunning{Polynomial affine model of gravity in three dimensions}
% if too long for running head

\author{Oscar Castillo-Felisola\thanksref{addr1,e1}
        \and
        Oscar Orellana\thanksref{addr2}
        \and
        Jos\'e Perdiguero\thanksref{addr1}
        \and
        Francisca Ram\'irez\thanksref{addr1}
        \and
        Aureliano Skirzewski\thanksref{addr3}
        \and
        Alfonso R. Zerwekh\thanksref{addr1}
}

%% \thankstext{t1}{Grants or other notes about the article that should go
%%   on the front page should be placed here. General acknowledgments
%%   should be placed at the end of the article.}
\thankstext{e1}{e-mail: o.castillo.felisola@protonmail.com}
%% \thankstext{e2}{e-mail: sauthor@example.com}

\authorrunning{O.~Castillo-Felisola, et. al.} % if too long for running head

\institute{\UTFSM\label{addr1} \and \UTFSMmat\label{addr2} \and \UdelaR\label{addr3}}

\date{Received: date / Accepted: date}

\maketitle

\begin{abstract}
The polynomial affine gravity is a model that is built up without the
explicit use of a metric tensor field. In this article we reformulate
the three-dimensional model and, given the decomposition of the affine
connection, we analyse the consistently truncated sectors. Using the
cosmological ansatz for the connection, we scan the cosmological
solutions on the truncated sectors. We discuss the emergence of
different kinds of metrics.
%% We reformulate the polynomial affine model of gravity in three
%% dimensions, analyse the consistently truncated sectors and find
%% solutions on these sectors.
\end{abstract}

\section{Introduction}
\label{sec:intro}
Our current understanding of fundamental physics accounts for four
different interactions, which split into two pillars. On the one hand,
the gravitational interaction is described by General Relativity,
which models the spacetime as a Riemannian manifold \((\Mi,g)\), i.e.
its geometrical properties are tied up to the metric tensor field. On
the other hand, the remaining three interactions are described by
gauge theories, whose fundamental field are connections under
transformations of the gauge group, but (covariant) vectors under the
group of diffeomorphisms.

The formulation of the gauge theories requires the choice of a
background manifold, on which the theories stand, and the standard
treatment does not consider back-reaction of the gauge theories to the
manifold. Nonetheless, in General Relativity the metric tensor field
defines the geometric properties of the spacetime, and also mediates
the gravitational interaction. Such dual role of the metric tensor
field originates significant differences between these pillars of
fundamental physics, being the most highlighted the quantisation
program. While the attempts to quantise General Relativity have eluded 
a consistent culmination, the remaining three interactions are
quantisable and renormalisable, raising what is known as the standard
model of particles.

In a quest to prevent the \emph{inconsistent} quantisation of the
gravitational interaction, a large number of generalisations of
General Relativity have been proposed. Inspired by the seminal work by
Palatini \cite{palatini19_deduz_invar_delle_equaz_gravit}, a large
amount of these generalisations modelled the spacetime by manifolds
whose connections is not the one of Levi-Civita, dubbed
\emph{metric-affine} models of gravity (see for example Ref.
\cite{hehl95_metric_affin_gauge_theor_gravit}). These models rely in the
metric as the field mediating the interaction. However, some have
attempted to model the gravitational interaction through affine
theories, where the mediating field is an affine (linear) connection
on the spacetime. The first proposals of affine gravity were
considered by Einstein, Eddington and Schrödinger
\cite{einstein23_zur_affin_feldt,einstein23_theor_affin_field,eddington23,schroedinger50_space},
but those models were left aside because their manipulation was
significantly more complex and offered insufficient novelty from the
phenomenological point of view. More contemporaneous affine models of
gravity were proposed by Kijowski and collaborators
\cite{kijowski78_new_variat_princ_gener_relat,ferraris81_gener_relat_is_gauge_type_theor,ferraris82_equiv_relat_theor_gravit,kijowski07_univer_affin_formul_gener_relat},
Poplawski
\cite{poplawski07_nonsy_purel_affin,poplawski07_unified_purel_affin_theor_gravit_elect,poplawski14_affin_theor_gravit},
Krasnov and collaborators
\cite{krasnov07_non_metric_gravit,krasnov08_non_metric_gravit_i,krasnov08_non_metric_gravit_ii,krasnov11_pure_connec_action_princ_gener_relat,delfino15_pure_connec_formal_gravit_lin,delfino15_pure_connec_formal_gravit_feyn},
and some of us
\cite{castillo-felisola15_polyn_model_purel_affin_gravit,castillo-felisola18_einst_gravit_from_polyn_affin_model}.

The analysis of gravitational models in lower dimensions, where the
number of parameters decreases in general, serves as a playground to
test methods that might be applied to the four-dimensional models, but
also for their applications in other branches of physics since vector
and tensor gauge theories can be interpreted as the high-temperature
limit of four-dimensional models \cite{weinberg76_under}. These
\emph{simplified} models stimulate the generation of new ideas and insights
into their four-dimensional counterparts.

The three-dimensional version of General Relativity was firstly
considered by Staruszkiewicz
\cite{staruszkiewicz63_gravit_theor_three_dimen_space}, where a
Schwarzschild-like solution was studied, and a relation between the
presence of massive point particles and conical singularities (which
modify the asymptotic behaviour of the spacetime) was found. The
interest for three-dimensional was left aside due to the lack of
propagating degrees of freedom, until in a series of papers Deser,
Jackiw and collaborators showed that adding a nontrivial, gauge
invariant, topological term to the three-dimensional Einstein--Hilbert
action resulted in a massive model for gravity
\cite{deser82_three_dimen_massiv_gauge_theor,deser82_topol_massiv_gauge_theor,deser84_three_dimen_cosmol_gravit,deser84_three_dimen_einst_gravit}.
The topological term added in this model, dubbed
Deser--Jackiw--Templeton, was the Chern--Simons term associated to the
four-dimensional \(\theta\)-term build from the (metric) Riemannian
curvature---also known as the Pontryagin density. A remarkable feature
of the Deser--Jackiw--Templeton model is that it was able to induce a
mass for the graviton without a Higgs mechanism and preserving the
(infinitesimal) gauge invariance.

Witten showed in Ref.
\cite{witten88_dimen_gravit_as_exact_solub_system}, that the
three-dimensional gravity modified by the Pontryagin Chern--Simons
Lagrangian is equivalent to a Yang--Mills theory, called Chern--Simons
gravity, and also that its perturbative expansion was
renormalisable.\footnote{An update on the original ideas in this paper can be found in
Ref. \cite{witten07_three_dimen_gravit_revis}.} Some years later, the first black hole solution
was found by Bañados, Teitelboim and Zanelli
\cite{banados92_black_hole_three_dimen_time}, disproving the
\emph{triviality} of classical three-dimensional gravity and revitalising
the interest in the search of exact solutions
\cite{garcia-diaz17_exact_solut_three_dimen_gravit} and further
development into their quantum aspects \cite{carlip98_quant}.
Contemporaneously, Mielke and Baekler generalised the topological
massive model of gravity to include torsion
\cite{mielke91_topol_gauge_model_gravit_with_torsion}, and extended
further to metric-affine gravity by Tresguerres
\cite{tresguerres92_topol_gravit_three_dimen_metric_affin_time}.

An interconnection between three-dimensional gravity with other
branches of physics was encountered by Dereli and Verçin in the
context of the continuum theory of lattice defects, through the
identification of the dislocation and disclination line density tensors
with torsion and curvature tensors, and the free-energy density with
the Lagrangian of the Deser--Jackiw--Templeton model of gravity
\cite{dereli91_gauge_model_amorp_solid_contain_defec_ii}.\footnote{For a more recent review, see Ref. \cite{lazar10_cartan_spiral_stair_physic_and}.} More
recently, methods of quantum field theory in curved spaces have been
applied to the analysis of properties of graphene, viewed as a
\emph{membrane} endowed with an metric induced by its embedding into
three-dimensional space
\cite{kerner12_inter_flexur_phonon_with_elect_graph,chaves14_optic_conduc_curved,pacheco14_graph_morph_elect_proper_from,castro-villarreal17_pseud_field_curved,oliveira17_signat_curved_qft_effec_optic,arrighi19_from_curved_spacet_to_spacet}.

From a mathematical point of view, the Deser--Jackiw--Templeton model
generalises the three-dimensional Einstein--Hilbert model by taking
into account global properties of the Riemannian spacetime. The
Mielke--Baekler model relaxes the Riemannian condition, by allowing
the spacetime to be modelled by a Riemann--Cartan manifold, while
in the Tresguerres model the spacetime possesses also non-metricity.
In all of these models, the metric plays a fundamental role in their 
formulation. However, the existence of contexts in which the notion of
metric is not helpful, e.g. in a phase space or a moduli space,
inspires the search of \emph{gravitational} models which are defined in
spaces where the notion of displacement and flow is defined, but not
necessarily a notion of length. These spaces are called \emph{affine
manifolds}, and they are the underlying structure supporting the
polynomial affine model of gravity.

\hspace*{-\parindent}%
\begin{minipage}[c]{.4\textwidth}
The classification of the affinely connected spaces is shown
schematically in the following commutative diagram. The most general
affine manifold might not posses a metric, and thus it is
characterised by a connection \((\Mi,\hat{\nabla})\). When an affine
manifold is equipped with a metric, dubbed metric-affine manifold, the 
connection can be decomposed into three contributions---the
Levi-Civita connection, the contorsion tensor and the deflection
tensor---, and the manifold is classified according to its curvature
\((\mathcal{R})\), torsion \((\mathcal{T})\) and non-metricity
\((\mathcal{Q})\). In the diagram, Riemannian geometries belong to the
sector denoted by \((\mathcal{R})\), Riemann--Cartan geometries to
\((\mathcal{R},\mathcal{T})\), Weizenböck geometries to
\((\mathcal{T})\), et cetera.
\end{minipage}
\begin{minipage}[c]{.6\textwidth}
\begin{center}
\begin{tikzcd}[row sep=scriptsize, column sep=scriptsize]
  & (\Mi,\hat{\nabla}) \ar[d, "g"] & & \\
  & (\mathcal{Q},\mathcal{R},\mathcal{T}) \arrow[dl] \arrow[rr] \arrow[dd] & & (\mathcal{Q},\mathcal{T}) \arrow[dl] \arrow[dd] \\
  (\mathcal{Q},\mathcal{R}) \arrow[rr, crossing over] \arrow[dd] & & (\mathcal{Q}) \\
  & (\mathcal{R},\mathcal{T}) \arrow[dl] \arrow[rr] & & (\mathcal{T}) \arrow[dl] \\
  (\mathcal{R}) \arrow[rr] & & (\text{Flat}) \arrow[from=uu, crossing over]\\
\end{tikzcd}
\end{center}
\end{minipage}

The use of an affine model allows to explore features that should be
attributed to the local invariance under coordinate transformations,
regardless of the metric structure on the manifold. In this context,
the polynomial affine model of gravity emerges naturally. Even though
the four-dimensional model is being analysed, the three-dimensional
version is expected to be easier to characterise, and eventually solve
some issues the affine models have encountered, such as the coupling
of matter.

In this paper we re-state the three-dimensional model of polynomial
affine gravity, firstly proposed in Ref.
\cite{castillo-felisola15_polyn_model_purel_affin_gravit}, analyse
their field equations and find explicit cosmological solutions. The
paper is organised as follows. Section \ref{sec:model} gives a brief
overview of the polynomial affine model of gravity in three
dimensions. Then, in Sec. \ref{sec:field_equations} the field
equations are derived, and some issues regarding their
truncation---i.e. restriction to sectors where only a subset of the
fields (irreducible components of the connection) are \emph{turned
  on}---are discussed in Sec. \ref{sec:truncations}. In Sec.
\ref{sec:scan} we scan the space of solutions which are compatible
with the cosmological principle. Some conclusions are drawn in the Sec.
\ref{sec:conclusions}. For completeness we include some appendices,
including a discussion of our notation in \ref{app:notation}.

\section{Building the model}
\label{sec:model}
The polynomial affine model of gravity was born as an attempt to build
up a theory with the affine connection as sole fundamental field
\cite{castillo-felisola15_polyn_model_purel_affin_gravit}. Formally, the
idea behind the polynomial affine model of gravity is that spacetime
is not a (pseudo-)Riemannian manifold, but an affinely connected
manifold \((\Mi,\hat{\nabla})\). The strategy is then to consider all
possible terms, allowed by the invariance under diffeomorphisms, as
part of the Lagrangian density. Note that a generic affine connection
is a reducible object under the group of diffeomorphisms, and
therefore we can decompose it.

In the absence of a metric tensor field, the affine connection
decomposes into irreducible components as follows
\begin{equation}
  \ct*{\mu}{\lambda}{\nu} 
  =
  \ct*{(\mu}{\lambda}{\nu)} + \ct*{[\mu}{\lambda}{\nu]}
  =
  \ct{\mu}{\lambda}{\nu} + \bt{\mu}{\lambda}{\nu} + \Ag_{[\mu}\delta^\lambda_{\nu]},
  \label{eq:conn_decomp}
\end{equation}
where \(\bt{}{}{}\) field is the traceless part of the torsion,
\(\Ag_\mu\) is the trace of the torsion, and \(\ct{\mu}{\lambda}{\nu}
\equiv \ct*{(\mu}{\lambda}{\nu)}\) is a renaming of the symmetric part
of the affine connection. All these elements transform as tensors
under diffeomorphisms, with the exception of the symmetric part of the
affine connection, which must be included in the action almost
exclusively through the covariant derivative.

In order to build an action, we consider the chart-induced basis of
the tangent and cotangent spaces, i.e. \(\set{\partial_\mu}\) and
\(\set{\de{x}^\mu}\), and the volume form defined as
\begin{equation}
  \de{V}^{\alpha\beta\gamma} = \de{x}^\alpha \wedge \de{x}^\beta \wedge \de{x}^\gamma.
  \label{eq:def-volume}
\end{equation}
An interesting type of connection are those \emph{compatible} with the
volume, i.e. \(\nabla(\de{V}) = 0\), which are said to be equiaffine.
Such compatibility ensures that the Ricci tensor field is symmetric,
and the trace of the curvature tensor vanishes
\cite{eisenhart27_non_rieman,schouten13_ricci,nomizu94_affin}. Although
equiaffinity is not demanded in the following formulation, in Sec.
\ref{sec:scan} we shall encounter that symmetric connections compatible
with the cosmological principle are necessarily equiaffine. 

Now, with the aid of the irreducible components of the connection and
the volume form, we can write down the most general Lagrangian, i.e. a
scalar density in three dimensions. A \emph{dimensional analysis} similar
to the one presented in Refs.
\cite{castillo-felisola18_einst_gravit_from_polyn_affin_model,castillo-felisola18_beyond_einstein,castillo-felisola18_cosmol,castillo-felisola20_emerg_metric_geodes_analy_cosmol}
shows that the most general action (up to boundary terms) is given
by\footnote{Note that this action is equivalent to the one introduced in
Ref. \cite{castillo-felisola15_polyn_model_purel_affin_gravit}.}
\begin{dmath}
  \label{eq:action}
  S = \int \de{V}^{\alpha\beta\gamma} \bigg(
  B_1 \, \A_\alpha \A_\mu \bt{\beta}{\mu}{\gamma}
  + B_2 \, \A_\alpha \Fg_{\beta\gamma}
  + B_3 \, \A_\alpha \nabla_\mu \bt{\beta}{\mu}{\gamma}
  + B_4 \, \bt{\alpha}{\mu}{\nu} \bt{\beta}{\nu}{\lambda} \bt{\gamma}{\lambda}{\mu}
  + B_5 \ri{\alpha\beta}{\mu}{\mu} \A_\gamma
  + B_6 \, \ri{\mu\alpha}{\mu}{\nu} \bt{\beta}{\nu}{\gamma}
  + B_7 \, \ct{\alpha}{\mu}{\mu} \partial_\beta \ct{\gamma}{\nu}{\nu}
  + B_8 \, \left( \ct{\alpha}{\mu}{\nu} \partial_\beta \ct{\gamma}{\nu}{\mu} +
    \frac{2}{3} \ct{\alpha}{\mu}{\nu} \ct{\beta}{\nu}{\lambda} \ct{\gamma}{\lambda}{\mu} \right)
  \bigg).
\end{dmath}
In the action, the covariant derivative and the curvature are defined with
respect to the symmetric connection, i.e. \(\nabla = \nabla^\Gamma\)
and \(\ri{}{}{}=\ri{}{\Gamma}{}\).

Among the features of the polynomial affine model of gravity we
count: (i) The fundamental field is a connection, like the other
fundamental interactions; (ii) The coupling constant are
dimensionless, which is desirable from the view point of Quantum Field
Theory, since the superficial degree of divergence vanishes; (iii) The
model seems to exhibit scale invariance; (iv) The number of possible
terms in the action is finite---we usually refer to this property as
the \emph{rigidity} of the model---, giving the impression that in the
hypothetical scenario of quantisation all the counter-terms have the
form of terms already present in the original action.

\section{Field equations}
\label{sec:field_equations}
We now focus in obtaining the field equations for the fields
\(\Gamma\), \(\bt{}{}{}\) and \(\A\), by varying the action in Eq.
\eqref{eq:action}. It is important to highlight that although the
absence of second-class constraints in Polynomial Affine Gravity has
not been proven yet, the structure of the action suggests that the
variational problem is well-posed, and therefore the field equations
below do not consider the existence of affine analogues of the
Gibbons--Hawking--York term.\footnote{Analysis of affine analogues to the Gibbons--Hawking--York term
can be found in
Refs. \cite{parattu16_bound_term_gravit_action_with_null_bound,krishnan17_robin_gravit,krishnan17_neuman_bound_term_gravit,lehner16_gravit_action_with_null_bound,hopfmueller17_gravit_degrees_freed_null_surfac,jubb17_bound_corner_terms_action_gener_relat}.}

Since the action contains up to first derivatives of the fields,
the field equations are obtained through the Euler-Lagrange equations,
\begin{equation}
  \begin{aligned}
    \partial_\mu \left(\frac{\partial \Lag}{\partial\left(\partial_\mu{\ct{\nu}{\lambda}{\rho}}\right)}\right)
    - \frac{\partial \Lag}{\partial {\ct{\nu}{\lambda}{\rho}}} & = 0,
    &
    \partial_\mu \left(\frac{\partial \Lag}{\partial\left(\partial_\mu{\bt{\nu}{\lambda}{\rho}}\right)}\right)
    - \frac{\partial \Lag}{\partial {\bt{\nu}{\lambda}{\rho}}} & = 0,
    &
    \partial_\mu \left(\frac{\partial \Lag}{\partial\left(\partial_\mu{ \Ag_{\nu}}\right)}\right) -
    \frac{\partial \Lag}{\partial { \Ag_{\nu}}} & = 0.
  \end{aligned}
  \label{eq:euler_lagrange} 
\end{equation}
We proceed utilising the formalism introduced by Kijowski in Ref.
\cite{kijowski78_new_variat_princ_gener_relat}, and following the steps
from Ref. \cite{castillo-felisola20_emerg_metric_geodes_analy_cosmol}.

It can be shown with ease that the Euler--Lagrange equations
\eqref{eq:euler_lagrange} can be rewritten as
\begin{equation}
  \begin{aligned}
    %\label{eq:FEQ_G}
    \nabla_\mu \PG{\mu\nu}{\lambda}{\rho} & = \frac{\partial^* \Lag}{\partial \ct{\nu}{\lambda}{\rho} },
    &
    % \label{eq:FEQ_B}
      \nabla_\mu \PG*{\mu\nu}{\lambda}{\rho} & = \frac{\partial \Lag}{\partial \bt{\nu}{\lambda}{\rho} },
    &
    % \label{eq:FEQ_A}
      \nabla_\mu \PG**{\mu\nu}{}{} & = \frac{\partial \Lag}{\partial \Ag_{\nu} },
  \end{aligned}
  \label{eq:FEQ_G}
\end{equation}
where the quantities \(\Pi_X\) are the canonical momentum associated
to the field \(X\)---which are tensor densities---, defined as
\begin{equation}
  \begin{aligned}
  \PG{\mu\nu}{\lambda}{\rho} & = \frac{ \partial \Lag }{ \partial \, (\partial_\mu \ct{\nu}{\lambda}{\rho}) },
  &
  \PG*{\mu\nu}{\lambda}{\rho} & = \frac{ \partial \Lag }{ \partial \, (\partial_\mu \bt{\nu}{\lambda}{\rho}) },
  &
  \PG**{\mu\nu}{}{} & = \frac{ \partial \Lag }{ \partial \, (\partial_\mu \Ag_{\nu}) },
  \end{aligned}
  \label{eq:canonical_momentum}
\end{equation}
and the asterisk on the right-hand side of the field equation for the
symmetric part of the connection in Eq. \eqref{eq:FEQ_G} denotes
the partial derivative with respect to the connection that is not
contained in the curvature tensor.

The field equations for the fields \(\A\), \(\bt{}{}{}\) and
\(\Gamma\) derived from the action in Eq. \eqref{eq:action} are
respectively:
\begin{dmath}
  2 B_1 \A_\alpha \bt{\nu}{\alpha}{\rho} + 2 B_2
  \Fg_{\nu\rho} + B_3 \nabla_\mu \bt{\nu}{\mu}{\rho} + B_5
  \ri{\nu\rho}{\mu}{\mu} = 0,
  \label{eq:feq_A}
\end{dmath}
\begin{dmath}
  2 B_1 \A_\nu \A_\rho - 2 B_3 \nabla_{(\nu} \A_{\rho)} + 3 B_4
  \bt{\nu}{\mu}{\sigma} \bt{\rho}{\sigma}{\mu} + 2 B_6
  \ri{\mu(\nu}{\mu}{\rho)} = 0,
  \label{eq:feq_B}
\end{dmath}
\begin{dmath}
  B_3 \A_\mu \bt{\rho}{\nu}{\sigma}
+ B_5 \left( \delta_\mu^\nu
    \Fg_{\rho\sigma} +  \delta_{[\rho}^\nu \Fg_{\sigma]\mu} \right)
+ B_6 \left( 2 \delta_{\mu}^{\nu} \nabla_{\tau}{ \bt{\rho}{\tau}{\sigma} }
    + \delta_{\rho}^{\nu} \nabla_{\tau}{ \bt{\sigma}{\tau}{\mu} }
    + \delta_{\sigma}^{\nu} \nabla_{\tau}{ \bt{\mu}{\tau}{\rho} }
    \right)
+ B_7 \left( \delta_\mu^\nu
    \ri{\rho\sigma}{\lambda}{\lambda} + \delta_{[\rho}^\nu 
    \ri{\sigma]\mu}{\lambda}{\lambda} \right)
+ B_ 8 \left( \ri{\rho\sigma}{\nu}{\mu} + \delta_{[\rho}^\nu
    \ri{\sigma]\lambda}{\lambda}{\mu} \right) = 0.
  \label{eq:feq_G}
\end{dmath}

Note that the field equation for the \(\bt{}{}{}\)-field, i.e. Eq.
\eqref{eq:feq_B}, is similar to the Ricci form of the Einstein field
equations. Particularly, it has been shown that the tensor
\(\bt{\nu}{\mu}{\sigma} \bt{\rho}{\sigma}{\mu}\) (when
non-degenerated) might be interpreted as a torsion-descendent metric
tensor field \cite{poplawski14_affin_theor_gravit}.

\section{Truncations}
\label{sec:truncations}
Unlike the analysis of the truncations in the four-dimensional
polynomial affine model of gravity
\cite{castillo-felisola18_einst_gravit_from_polyn_affin_model,castillo-felisola18_cosmol,castillo-felisola20_emerg_metric_geodes_analy_cosmol},
the presence of the Chern--Simons terms in \eqref{eq:action} allows to
switch-off certain fields at the level of the action, and not just at
the level of the field equations. In order to distinguish these two
possible \emph{truncations} of the model, we refer to them as off-shell and
on-shell limits. Below we list six possible truncations of the model.

\subsection{Torsion-free limit}
\label{sec:orgce1601d}

On the one hand, taking the off-shell torsion-free limit yields an
effective action which is the sum of two Chern-Simons terms, 
\begin{equation}
  S_{\text{off}}[\Gamma] = \int \de{V}^{\alpha\beta\gamma} \left[ 
  B_7 \, \ct{\alpha}{\mu}{\mu} \partial_\beta \ct{\gamma}{\nu}{\nu}
  + B_8 \, \left( \ct{\alpha}{\mu}{\nu} \partial_\beta \ct{\gamma}{\nu}{\mu} +
    \frac{2}{3} \ct{\alpha}{\mu}{\nu} \ct{\beta}{\nu}{\lambda} \ct{\gamma}{\lambda}{\mu} \right)
  \right],
  \label{eq:offshell_action_torsion_free}
\end{equation}
whose field equations are
\begin{equation*}
  B_7 \left( \delta_\mu^\nu \ri{\rho\sigma}{\lambda}{\lambda} + \delta_{[\rho}^\nu \ri{\sigma]\mu}{\lambda}{\lambda} \right) + B_ 8 \left( \ri{\rho\sigma}{\nu}{\mu} + \delta_{[\rho}^\nu \ri{\sigma]\lambda}{\lambda}{\mu} \right) = 0.
  % \left( B_7 \, \partial_\alpha \ct{\beta}{\sigma}{\sigma} \delta_\lambda^{(\rho} + B_8 \ri{\alpha\beta}{(\rho}{\lambda} \right)  \de{V}^{\nu)\alpha\beta} = 0.
\end{equation*}
Note that the contribution to the field equations of the term with
coefficient \(B_7\) vanishes identically for equiaffine
connections,\footnote{For equiaffine connections the skew-symmetric part of the Ricci
tensor field vanishes. From the first (or algebraic)  Bianchi
identity, it follows that the trace of the curvature vanishes.
Therefore, \(\ri{\mu\nu}{\alpha}{\alpha} = \partial_{[\mu}
\ct{\nu]}{\alpha}{\alpha} = 0\). In the remaining of the article we
shall ignore the contribution of this term to the field equations,
however, we keep the term in the action because it might contribute to
topological quantities.} and thus the field equations for the symmetric
connection are
\begin{equation}
  B_ 8 \left( \ri{\rho\sigma}{\nu}{\mu} + \delta_{[\rho}^\nu \ri{\sigma]\lambda}{\lambda}{\mu} \right) = 0.
  % B_8 \, \ri{\alpha\beta}{(\rho}{\lambda} \de{V}^{\nu)\alpha\beta} = 0.
  \label{eq:offshell_feqs_torsion_free}
\end{equation}
These field equations require that the connection is projectively
Weyl-flat.\footnote{Weyl introduced two different notions of curvature, both of
therm are referred as Weyl's tensors, and sometimes they are called
\emph{projective} and \emph{conformal} Weyl tensor \cite{weyl21_zur_infin}. As
physicists we are custom to the conformal Weyl tensor, which might is
the curvature without traces (with respect to the metric). The
projective Weyl tensor might be defined without requiring a metric,
and it is invariant under the projective transformations of the
connection, \(\ct,{\mu}{\lambda}{\nu} = \ct{\mu}{\lambda}{\nu} +
\delta^\lambda_\mu V_\nu + \delta^\lambda_\nu V_\mu\). See Refs.
\cite{eisenhart27_non_rieman,nomizu94_affin,luebbe13_note_coinc_projec_confor_weyl_tensor}.}

On the other hand, when one takes the on-shell torsion-free sector,
there are subsidiary equations,
\begin{equation*}
  \ri{\nu\rho}{\mu}{\mu} = 0 \text{ and } \ri{\mu(\nu}{\mu}{\rho)} = 0,
\end{equation*}
the first is satisfied when the connection is equiaffine, while the
second (Ricci-flatness) restricts the solutions of the system to be
flat connections.

\subsection{Vectorial torsion}
\label{sec:org6cd961d}

The restriction to sole vectorial torsion in the off-shell limit
yields a Chern--Simons-like action, whose field equations are
\begin{equation}
  \Fg_{\nu\rho} = 0,
  \label{eq:offshell_feq_A}
\end{equation}
while the on-shell truncation raises the auxiliary condition
\begin{align}
  2 B_1 \A_\nu \A_\rho - 2 B_3 \partial_{(\nu} \A_{\rho)} & = 0.
  \label{eq:onshell_feq_A}
\end{align}

The system of equations \eqref{eq:offshell_feq_A} and
\eqref{eq:onshell_feq_A} can be solved in general. First, equation
\eqref{eq:offshell_feq_A} implies that locally the field \(\A\) is an
exact \(1\)-form, \(\A = \de{\phi}\). Secondly, equation
\eqref{eq:onshell_feq_A} can be written in the form
\begin{equation*}
  \partial_\nu \partial_\rho f = 0, \text{ with } f = e^{ - \frac{B_1}{B_3} \phi }.
\end{equation*}
The above equation implies that \(f\) is a linear function of the
coordinates, and allows to solve for \(\phi\) and therefore \(\A\),
\begin{equation*}
  \phi(x) = - \frac{B_3}{B_1} \ln \left( D + C_\mu x^\mu \right)
  \text{ and }
  \A_\nu(x) = - \frac{B_3}{B_1} \frac{C_\nu}{D + C_\mu x^\mu}.
\end{equation*}

Note that the field equations \eqref{eq:offshell_feq_A}, implies that
the \(\A\)-field inherits a gauge redundancy. This gauge redundancy
disappears with the existence of the \(\bt{}{}{}\)-field, even if it
vanishes, i.e. \(\bt{}{}{}\) breaks the gauge redundancy of \(\A\).

\subsection{Trace-less torsion}
\label{sec:orgcd90778}

Interestingly, the off-shell truncation to trace-less torsion leaves
an effective action whose sole term is that with coefficient \(B_4\),
which provides no dynamics to the \(\bt{}{}{}\)-field. The field
equations are
\begin{equation}
  \bt{\nu}{\mu}{\sigma} \bt{\rho}{\sigma}{\mu} = 0.
  \label{eq:offshell_feq_B}
\end{equation}

The on-shell truncation yields the subsidiary equations,
\begin{equation}
  \partial_{\mu} \bt{\rho}{\mu}{\sigma} = 0 \text{ and } \partial_{[\mu} \bt{\rho}{\nu}{\sigma]} = 0.
  \label{eq:onshell_feq_B}
\end{equation}
Note that these equations are equivalent,\footnote{Contracting the later with \(\epsilon^{\mu\rho\sigma}\) yields
the former.} and their equivalency
can be extended to the covariant version of the equations.

Equation \eqref{eq:offshell_feq_B} can be rewritten in terms of the
quasi-Hodge dual \cite{vaz16_introd_cliff_algeb_spinor},
\begin{equation}
  T^{\mu\alpha} = \frac{1}{2} \bt{\beta}{\mu}{\gamma} \epsilon^{\alpha\beta\gamma},
  \label{eq:b_dual}
\end{equation}
giving a cofactor equation for the tensor \(T\). Hence,
\(T^{\mu\alpha} = \rho(x) V^\mu V^\alpha\) is the general solution
where \(\rho\) is a scalar density, and therefore 
\begin{equation*}
  \bt{\nu}{\mu}{\lambda} = \rho(x) V^\mu V^\sigma \epsilon_{\sigma\nu\lambda}.
\end{equation*}

The structure of our tensor \(T^{\mu\alpha}\) is equivalent to that of
the energy-momentum tensor for cold matter (i.e. \emph{dust}), where
\(\rho\) is the energy density and \(V^\mu\) represents the velocity
of the matter distribution. Similarly, the Eq. \eqref{eq:onshell_feq_B}
is a generalisation of the continuity equation and  energy-momentum
conservation, 
\begin{dmath*}
  0 = \partial_\mu \bt{\nu}{\mu}{\lambda}
  = \left( (\partial_\mu \rho) V^\mu V^\sigma + \rho (\partial_\mu V^\mu) V^\sigma + \rho V^\mu (\partial_\mu V^\sigma) \right) \epsilon_{\sigma\nu\lambda}
  % = \left[ \nabla_V (\rho V) + (\nabla \cdot V) \rho V \right]^\sigma \epsilon_{\sigma\nu\lambda}
  = \left[ \rho \nabla_{V} V + (\nabla \cdot (\rho V)) V \right]^\sigma \epsilon_{\sigma\nu\lambda}.
\end{dmath*}
The last line is the equation for a self-parallel vector, written in
with a non-canonical affine parameter. The expression in brackets
corresponds to the divergence of the tensor \(T\), i.e. \(\partial_\mu
T^{\mu\sigma}\), which represents---in our analogue with the cold
matter energy-momentum tensor---the momentum conservation. However, in
General Relativity the each term in the bracket vanishes
independently, corresponding to the geodesic and continuity equations
respectively. 

\subsection{Symmetric connection with vectorial torsion}
\label{sec:org9ca832c}

The \emph{off-shell} limit, yielding an effective action that is the sum of the
three Chern-Simons terms (whose coupling constants are \(B_2\),
\(B_7\) and \(B_8\)) plus an interaction coming from the term whose
coupling constant is \(B_5\),
\begin{dgroup}
  \begin{dmath}
    0 = 2 B_2 \Fg_{\nu\rho} + B_5 \ri{\nu\rho}{\mu}{\mu},
    \label{eq:offshell_feq_GA1}
  \end{dmath}
  \begin{dmath}
    0 =
    B_5 \left( \delta_\mu^\nu \Fg_{\rho\sigma} + \delta_{[\rho}^\nu
      \Fg_{\sigma]\mu} \right) + B_7 \left( \delta_\mu^\nu
      \ri{\rho\sigma}{\lambda}{\lambda} + \delta_{[\rho}^\nu
      \ri{\sigma]\mu}{\lambda}{\lambda} \right) + B_ 8 \left(
      \ri{\rho\sigma}{\nu}{\mu} + \delta_{[\rho}^\nu
      \ri{\sigma]\lambda}{\lambda}{\mu} \right).
    \label{eq:offshell_feq_GA2}
  \end{dmath}
  \label{eq:offshell_feq_GA}
\end{dgroup}
Since Eq. \eqref{eq:offshell_feq_GA1} relates the field strength \(\Fg\)
with the trace of the curvature tensor, Eq. \eqref{eq:offshell_feq_GA2}
become an equation for just curvature objects. In particular, if the
coupling constants satisfy
\begin{equation}
  \left( \frac{2 B_2 B_7 - B_5^2}{2 B_2 B_8} \right) = \frac{1}{4},
  \label{eq:coefficient_weyl_ga}
\end{equation}
equation \eqref{eq:offshell_feq_GA2} coincides with Weyl's projective
curvature tensor field, which in three dimensions is
\begin{dmath*}
  \mathcal{W}_{\mu \nu}\,^{\lambda}\,_{\rho} = \mathcal{R}_{\mu \nu}\,^{\lambda}\,_{\rho}
  - \frac{1}{4} \delta^{\lambda}_{\rho} \ri{\mu\nu}{\sigma}{\sigma}
  - \frac{1}{2} \left( \ri{\sigma\nu}{\sigma}{\rho} \delta^{\lambda}_{\mu}
    - \ri{\sigma \mu}{\sigma}{\rho} \delta^{\lambda}_{\nu} \right)
  - \frac{1}{8} \left( \delta^{\lambda}_{\mu} \ri{\nu \rho}{\sigma}{\sigma}
    - \delta^{\lambda}_{\nu} \ri{\mu \rho}{\sigma}{\sigma} \right).
\end{dmath*}
Therefore, in this sector field equations describe a symmetric
projectively-flat connection. A projectively-flat connection is
\emph{locally} written as
\begin{equation*}
  \Gamma_{\nu}\,^{\lambda}\,_{\rho} = \delta_{\nu}^{\lambda} \psi_{\rho} + \delta_{\rho}^{\lambda} \psi_{\nu},
\end{equation*}
where \(\psi_\mu\) is a generic (differentiable) vector field. From
Eq. \eqref{eq:offshell_feq_GA1}, the \(\A\)-field is proportional to
\(\psi_\mu\), 
\begin{equation*}
  \A_\mu = - \frac{2 B_5}{B_2} \psi_\mu.
\end{equation*}

Note that if the coefficients do not satisfy Eq.
\eqref{eq:coefficient_weyl_ga}, the two equations in
\eqref{eq:offshell_feq_GA} are incompatible unless the trace of the
curvature and the field strength vanish independently, i.e.
\begin{equation*}
  \psi_\mu = \partial_\mu \alpha \text{ and } \A_\mu = \partial_\mu \beta.
\end{equation*}

In the on-shell limit the system of field equations are enriched by the
additional equation,
\begin{equation}
  B_1 \A_\nu \A_\rho - B_3 \nabla_{(\nu} \A_{\rho)} + B_6 \ri{\mu(\nu}{\mu}{\rho)} = 0,
  \label{eq:onshell_feq_GA}
  % \left( - B_1 \, \A_\alpha \A_\lambda + B_3 \, \nabla_\lambda \A_\alpha - B_6 \, \ri{\mu\alpha}{\mu}{\lambda} \right) \de{V}^{\nu\rho\alpha} = 0.
  % \label{eq:onshell_feq_vector_torsion}
\end{equation}
Note that the nontrivial part of the equation comes from the symmetric
part, which is an Einstein-like equation.

\subsection{Symmetric connection with trace-less torsion}
\label{sec:orgc71e8f3}

The off-shell restriction to this sector involves the terms in the
action, in Eq. \eqref{eq:action}, with coefficients from \(B_4\),
\(B_6\), \(B_7\) and \(B_8\). The field equations on this off-shell
limit are,
\begin{dmath}
  3 B_4 \bt{\nu}{\mu}{\sigma} \bt{\rho}{\sigma}{\mu} + 2 B_6 \ri{\mu(\nu}{\mu}{\rho)} = 0,
  \label{eq:offshell_feq_gb1}
\end{dmath}
\begin{dmath}
    B_6 \left( 2 \delta_{\mu}^{\nu} \nabla_{\tau}{ \bt{\rho}{\tau}{\sigma} }
    + \delta_{\rho}^{\nu} \nabla_{\tau}{ \bt{\sigma}{\tau}{\mu} }
    + \delta_{\sigma}^{\nu} \nabla_{\tau}{ \bt{\mu}{\tau}{\rho} }
    \right)
  + B_7 \left( \delta_\mu^\nu
      \ri{\rho\sigma}{\lambda}{\lambda} + \delta_{[\rho}^\nu
      \ri{\sigma]\mu}{\lambda}{\lambda} \right) +
    B_ 8 \left( \ri{\rho\sigma}{\nu}{\mu} + \delta_{[\rho}^\nu
    \ri{\sigma]\lambda}{\lambda}{\mu} \right) = 0.
  \label{eq:offshell_feq_gb2}
\end{dmath}

In the on-shell limit, there is an extra subsidiary condition, 
\begin{equation}
  B_3 \nabla_\mu \bt{\nu}{\mu}{\rho} + B_5 \ri{\nu\rho}{\mu}{\mu} = 0.
  % B_3 \, \nabla_\mu \bt{\alpha}{\mu}{\beta} \de{V}^{\nu\alpha\beta} = 0
  \label{eq:onshell_feq_gb}
\end{equation}
Note that for volume-preserving connections, Eq.
\eqref{eq:onshell_feq_gb} is nothing but the continuity equation of the
\(T\)-tensor (quasi-Hodge dual of the \(\bt{}{}{}\)-field) which might
be interpreted as a conserved ``energy-momentum tensor'', or in case
of being non-degenerated it admits the interpretation of a compatible
\emph{inverse metric} tensor field. Similarly, the
term accompanying the coupling constant \(B_7\) in Eq.
\eqref{eq:offshell_feq_gb2} vanishes.

Equation \eqref{eq:offshell_feq_gb1} is an Einstein-like equation, where
the term \(\bt{\nu}{\mu}{\sigma} \bt{\rho}{\sigma}{\mu}\) behaves like
a torsion-descendent metric.\footnote{Along the paper we might refer to this torsion-descendent
metric tensor field as the \emph{Poplawski metric}, due to the analysis of
such geometrical object by N.~Poplawski in
Ref.~\cite{poplawski14_affin_theor_gravit}.}

\subsection{The torsion sector}
\label{sec:org47fb848}

The off-shell limit to the torsion sector of our model is given
by restricting the action \eqref{eq:action} to the terms with
coefficients from \(B_1\) to \(B_4\). The field equations in this
limit are
\begin{dgroup}
  \begin{dmath}
    2 B_1 \A_\alpha \bt{\nu}{\alpha}{\rho} + 2 B_2 \Fg_{\nu\rho} + B_3 \partial_\mu \bt{\nu}{\mu}{\rho} = 0,
  \end{dmath}
  \begin{dmath}
    2 B_1 \A_\nu \A_\rho - 2 B_3 \partial_{(\nu} \A_{\rho)} + 3 B_4 \bt{\nu}{\mu}{\sigma} \bt{\rho}{\sigma}{\mu} = 0.
  \end{dmath}
  \label{eq:offshell_feq_AB}
\end{dgroup}

However, the on-shell restriction to the torsion sector yields an
extra condition,
\begin{dmath}
  B_3 \A_\mu \bt{\rho}{\nu}{\sigma}
+ B_5 \left( \delta_\mu^\nu
    \Fg_{\rho\sigma} +  \delta_{[\rho}^\nu \Fg_{\sigma]\mu} \right) +
  B_6 \left( 2 \delta_{\mu}^{\nu} \partial_{\tau}{ \bt{\rho}{\tau}{\sigma} }
    + \delta_{\rho}^{\nu} \partial_{\tau}{ \bt{\sigma}{\tau}{\mu} }
    + \delta_{\sigma}^{\nu} \partial_{\tau}{ \bt{\mu}{\tau}{\rho} }
    \right) = 0.
  \label{eq:onshell_feq_AB}
\end{dmath}
\section{A scan of cosmological solutions}
\label{sec:scan}
\subsection{Ansatz for the connection}
\label{sec:org6fc7c1b}

In order to develop further aspects of the three-dimensional
polynomial affine model of gravity, we have to provide an ansatz for
the connection. The ansatz are found by solving the equations derived
from the vanishing Lie derivative of the connection, i.e. 
\begin{dmath}
  \pounds_V \ct*{\mu}{\lambda}{\nu} =
  V^\sigma \partial_\sigma \ct*{\mu}{\lambda}{\nu}
  - \ct*{\mu}{\sigma}{\nu} \partial_\sigma V^\lambda
  + \ct*{\sigma}{\lambda}{\nu} \partial_\mu V^\sigma
  + \ct*{\mu}{\lambda}{\sigma} \partial_\nu V^\sigma
  + \frac{\partial^2 V^{\lambda}}{\partial x^{\mu} \partial x^\nu }
  = \hat{\nabla}_\mu \hat{\nabla}_\nu V^\lambda
  + \ri*{\rho\mu}{\lambda}{\nu} V^\rho
  - \hat{\nabla}_\mu \left( \tt{\nu}{\lambda}{\rho} V^\rho  \right)
  = 0,
  \label{eq:lie_der_connection}
\end{dmath}
where \(V\) represents a vector associated to the generators of the
symmetry group, i.e. each \(V\) defines a symmetry flow.

Two physically interesting cases are the \emph{isotropic} connection, which
is required to analyse spherical configurations, e.g. black hole
solutions, and the \emph{isotropic and homogeneous} connection, that is
compatible with the cosmological principle, and therefore required to
build cosmological models.

The vector fields generating the described symmetry groups are
\begin{equation}
  \begin{aligned}
    J & = %\partial_\varphi = 
    \begin{pmatrix}
      0 & 0 & 1
    \end{pmatrix},
    &%\\
    X & = %\partial_x =
    \sqrt{1 - \kappa r^2} 
    \begin{pmatrix}
      0 & \cos \varphi & - \frac{1}{r} \sin \varphi
    \end{pmatrix},
    &%\\
    Y & = %\partial_y =
    \sqrt{1 - \kappa r^2} 
    \begin{pmatrix}
      0 & \sin \varphi &  \frac{1}{r} \cos \varphi
    \end{pmatrix},
  \end{aligned}
  \label{eq:Killing_vectors}
\end{equation}
where \(J\) is associated with the sole angular momentum defining the
isotropy, while \(X\) and \(Y\) are associated with the translations
defining the homogeneity.

\subsubsection{Isotropic connection}
\label{sec:orgd964f16}

Since the components of the vector field \(J\) are constants, the Lie
derivative of the connection along its flow is equal to that of a
\(\binom{1}{2}\)-tensor. The vanishing Lie derivative along the vector
field \(J\) yields
\begin{equation}
  \pounds_J \ct*{\mu}{\lambda}{\nu} = \partial_\varphi \ct*{\mu}{\lambda}{\nu} = 0,
  \label{eq:isotropic_connection}
\end{equation}
i.e. this symmetry does not relate different components of the
connection, but its dependence on the coordinates. 

At this stage, without mentioning the field equations for the
connection, if one would try to find an analogue of the Schwarzschild
solution, there are \emph{twenty seven} functions to be determined.

\subsubsection{Isotropic and homogeneous connection}
\label{sec:org2063e5b}

Now, we require that the restricted components of the connection in
Eq. \eqref{eq:isotropic_connection}, are symmetric with respect to the
vector fields \(X\) and \(Y\).\footnote{Note that due to the isotropy, once calculating both Lie
derivatives is redundant, therefore we need to calculate and solve
only one of them.} The procedure to solve the
equations \(\pounds_X \ct{\mu}{\lambda}{\nu} = 0\) is tedious but
straightforward (see for example Refs.
\cite{castillo-felisola18_beyond_einstein,castillo-felisola18_cosmol}),
hence we only show the results below. This time, we split the affine
connection into its irreducible components,
\begin{equation}
  \begin{aligned}
    \ct{t}{t}{t} & = j(t),
    &
    \ct{i}{t}{j} & = g(t) S_{ij},
    \\
    \ct{i}{k}{j} & = \gamma_i{}^k{}_j,
    & \ct{t}{i}{j} & = \ct{j}{i}{t} = h(t) \delta^i_j + f(t) S^{ik} \epsilon_{kj} \frac{r}{\sqrt{1 - \kappa r^2}}, 
  \end{aligned} 
  \label{eq:homotropic_connection} 
\end{equation}
where \(f\), \(g\), \(h\) and \(j\) are functions of time, while \(S_{ij}\) and
\(\gamma_{i}{}^{j}{}_{k}\) are the two-dimensional rank two symmetric
tensor and connection compatible with isotropy and homogeneity,\footnote{It is worth mentioning that the general two-dimensional
isotropic and homogeneous (covariant and contravariant) tensors
posses a nonvanishing skew-symmetric component. This off-diagonal
component allows us to write the three-dimensional torsion full
connection in the same form of Eq. \eqref{eq:homotropic_connection} but
where the matrices \(S\) and \(S^{-1}\) are no longer symmetric.}
defined by
\begin{equation*}
  S_{ij} =
  \begin{pmatrix} 
    \frac{1}{1 - \kappa r^2} & 0 \\
    0 & r^2
  \end{pmatrix},
\end{equation*}
and 
\begin{align*}
  \gamma_{r}{}^{r}{}_{r} & = \frac{\kappa r}{1 - \kappa r^2},
  &
  \gamma_{\varphi}{}^{r}{}_{\varphi} & = - r ( 1- \kappa r^2 ),
  &
  \gamma_{r}{}^{\varphi}{}_{\varphi} & = \frac{1}{r},
  &
    \gamma_{\varphi}{}^{\varphi}{}_{r} & = \frac{1}{r}.
\end{align*}
It is important to highlight the existence of an unexpected function
in the components of the affine connection, to know, the
\(f\)-function, which might be introduced solely in the
three-dimensional case. Furthermore, as shown in
\ref{sec:t_reparam}, the function \(j\) can be set to zero by a
reparametrisation of the time coordinate.

The nonvanishing components of the \(\bt{}{}{}\)-field are 
\begin{equation}
  \begin{aligned}
    \bt{\varphi}{t}{r} = - \bt{r}{t}{\varphi} & = \xi(t) \frac{r}{\sqrt{1 - \kappa r^2}},
    &
    \bt{t}{r}{\varphi} = - \bt{\varphi}{r}{t} & = \psi(t) r \sqrt{1 - \kappa r^2},
    &
    \bt{r}{\varphi}{t} = - \bt{t}{\varphi}{r} & = \frac{\psi(t)}{r \sqrt{1 - \kappa r^2}},
  \end{aligned}
  \label{eq:homotropic_bt}
\end{equation}
while the nonvanishing component of the \(\A\)-field is \(\A_t = \eta(t)\).

\subsection{Curvature of the symmetric connection}
\label{sec:org490165e}

From Eq. \eqref{eq:homotropic_connection}, the components of curvature
are calculated, yielding
\begin{equation}
  \begin{split}
    \ri{t i}{t}{j} & = - \ri{i t}{t}{j} = \left( \dot{g} - g h \right) S_{ij} - f g \frac{r}{\sqrt{1 - \kappa r^2}} \epsilon_{ij},
    \\
    \ri{i j}{t}{t} & = 2 f g \frac{r}{\sqrt{1 - \kappa r^2}} \epsilon_{ij},
    \\
    \ri{t i}{j}{t} & = - \ri{i t}{j}{t} = \left( \dot{h} + h^2 - f^2 \right) \delta_i^j + \left( \dot{f} + 2 f h \right) \frac{r}{\sqrt{1 - \kappa r^2}} S^{jk} \epsilon_{ki},
    \\
    \ri{i j}{k}{l} & = 2 (g h + \kappa) S_{l[j} \delta_{i]}^k - f g \frac{r}{\sqrt{1 - \kappa r^2}} \epsilon_{ij} \delta_l^k.
  \end{split}
  \label{eq:homotropic_curvature}
\end{equation}
It follows that the trace of the curvature vanishes,
\(\ri{\mu\nu}{\sigma}{\sigma} = 0\), and therefore 
our connection is \emph{equiaffine}. Hence, the Ricci tensor field is
symmetric. 

The nonvanishing components of the Ricci tensor field are,
\begin{equation}
  \begin{aligned}
    \ri{tt}{}{} & = - 2 \left( \dot{h} + h^2 - f^2 \right),
    &
    \ri{ij}{}{} & = \left( \dot{g} + \kappa \right) S_{ij}.
  \end{aligned}
  \label{eq:homotropic_ricci}
\end{equation}

The three-dimensional Weyl projective curvature for an equiaffine
connection is then,
\begin{equation}
  \mathcal{W}_{\mu\nu}{}^\lambda{}_\rho = \ri{\mu\nu}{\lambda}{\rho} - \frac{1}{2} \left( \ri{\nu\rho}{}{} \delta_\mu^\lambda - \ri{\mu\rho}{}{} \delta_\nu^\lambda \right),
  \label{eq:weyl_curvature}
\end{equation}

Note that since in general three-dimensional connections are not
projectively flat, the curvature cannot be resolved in terms of the
Ricci tensor, unlike what it is expected from General Relativity.

Before turn over solving the differential equations, there are some
interesting remarks:
\begin{itemize}
\item The \(f^2\)-function is a \emph{parameter} for the system of ordinary
differential equations for isotropic and homogeneous Ricci-flat
spaces.
\item Isotropic and homogeneous flat spaces require that either \(f\) or
\(g\) vanishes.
\item The dynamical equations for \(f\) and \(g\) are \emph{always} integrable
in terms of \(h\) and \(\kappa\).
\item The dynamics of \(h\) is determined by a Riccati ordinary
differential equation, which is the same for both flat and
Ricci-flat spaces.
\end{itemize}
Since the Riccati equation is not solvable in general, we may
characterise the solutions of the system of ordinary differential
equations by the partial solutions of the Riccati equation.

\subsection{Space of solutions}
\label{sec:orgb105cee}

The substitution of the cosmological ansatz for the connection into
the field equations in Eqs. \eqref{eq:feq_A}, \eqref{eq:feq_B} and
\eqref{eq:feq_G}, yields the system
\begin{dgroup}[noalign]
  \begin{dmath}
    2 B_8 g f - B_3 \xi \eta = 0,
    \label{eq:oo1}
  \end{dmath}
  \begin{dmath}
    B_6 (2 g \psi + \dot{\xi}) - B_8 f g = 0,
    \label{eq:oo2}
  \end{dmath}
  \begin{dmath}
    B_8 (2 g h + \kappa - \dot{g}) = 0,
    \label{eq:oo3}
  \end{dmath}
  \begin{dmath}
    B_3 \eta \psi + B_8 (2 h f + \dot{f}) = 0,
    \label{eq:oo4}
  \end{dmath}
  \begin{dmath}
    B_1 \eta^2 - B_3 \dot{\eta} + 3 B_4 \psi^2 - 2 B_6 (\dot{h} + h^2
    - f^2) = 0,
    \label{eq:oo5}
  \end{dmath}
  \begin{dmath}
    B_3 g \eta - 3 B_4 \psi \xi + B_6 (\kappa + \dot{g}) = 0,
    \label{eq:oo6}
  \end{dmath}
  \begin{dmath}
    2 B_1 \eta \xi + B_3 (2 g \psi + \dot{\xi}) = 0.
    \label{eq:oo7}
  \end{dmath}
  \label{eq:cosmological_feqs}
\end{dgroup}

Note that the Eq. \eqref{eq:oo1} is algebraic, from Eqs. \eqref{eq:oo2}
and \eqref{eq:oo7} an algebraic relation is obtained after eliminating
the expression \(2 g \psi + \dot{\xi}\), and similarly from Eqs.
\eqref{eq:oo3} and \eqref{eq:oo6} after eliminating \(\dot{g}\). These
expressions are
\begin{dgroup}[noalign]
  \begin{dmath}
    2 B_8 g f - B_3 \xi \eta = 0,
    \label{eq:ooa}
  \end{dmath}
  \begin{dmath}
    \frac{B_8}{B_6} g f + 2 \frac{B_1}{B_3} \xi \eta = 0,
    \label{eq:oob}
  \end{dmath}
  \begin{dmath}
    3 \frac{B_4}{B_6} \psi \xi - 2 g h - \frac{B_3}{B_6} g \eta = 2
    \kappa.
    \label{eq:ooc}
  \end{dmath}
  \label{eq:oo_eqs}
\end{dgroup}
Equations \eqref{eq:ooa} and \eqref{eq:oob} can be seen as a system of
equations for the variables \(f\) and \(\eta\) as functions of \(g\)
and \(\xi\), but there independence is dictated by the determinant of
the system coefficients,
\begin{equation}
  \Omega = B_8 \left[ 4 \frac{B_1}{B_3} + \frac{B_3}{B_6} \right] g \xi.
  \label{eq:oo_omega}
\end{equation}

\tikzset{treenode/.style = {shape=rectangle, rounded corners, draw, align=center, top color=white, bottom color=blue!10}, root/.style = {treenode, font=\Large, bottom color=red!10}, exact/.style = {treenode, bottom color=green!30}, inexact/.style = {treenode, bottom color=yellow!30}}

We explore the space of cosmological solutions considering all
possible cases, which can be represented by \emph{decision trees}. On the
one hand, when the determinant in Eq. \eqref{eq:oo_omega} is
nonvanishing, it follows that \(f = \eta = 0\), and also that both
\(g\) and \(\xi\) are nonvanishing. The field equations allow to
parametrise \(\psi = \psi(g,h,\xi)\), and thus the final system
reduces to a set of three equations and three unknowns.

\begin{center}
\includegraphics[scale=1]{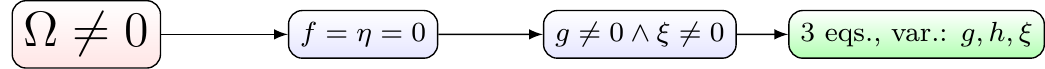}
\end{center}

On the other hand, when the determinant \(\Omega\) vanishes, following
the same approach, the decision tree has more branches.\footnote{Note that in the decision tree we do not include branches for
the cases where the parameters of the model are constrained, i.e. the
case \(B_8 = 0\) and \(B_3^2 + 4 B_1 B_6 = 0\).}

\begin{center}
  \includegraphics[width=.9\textwidth]{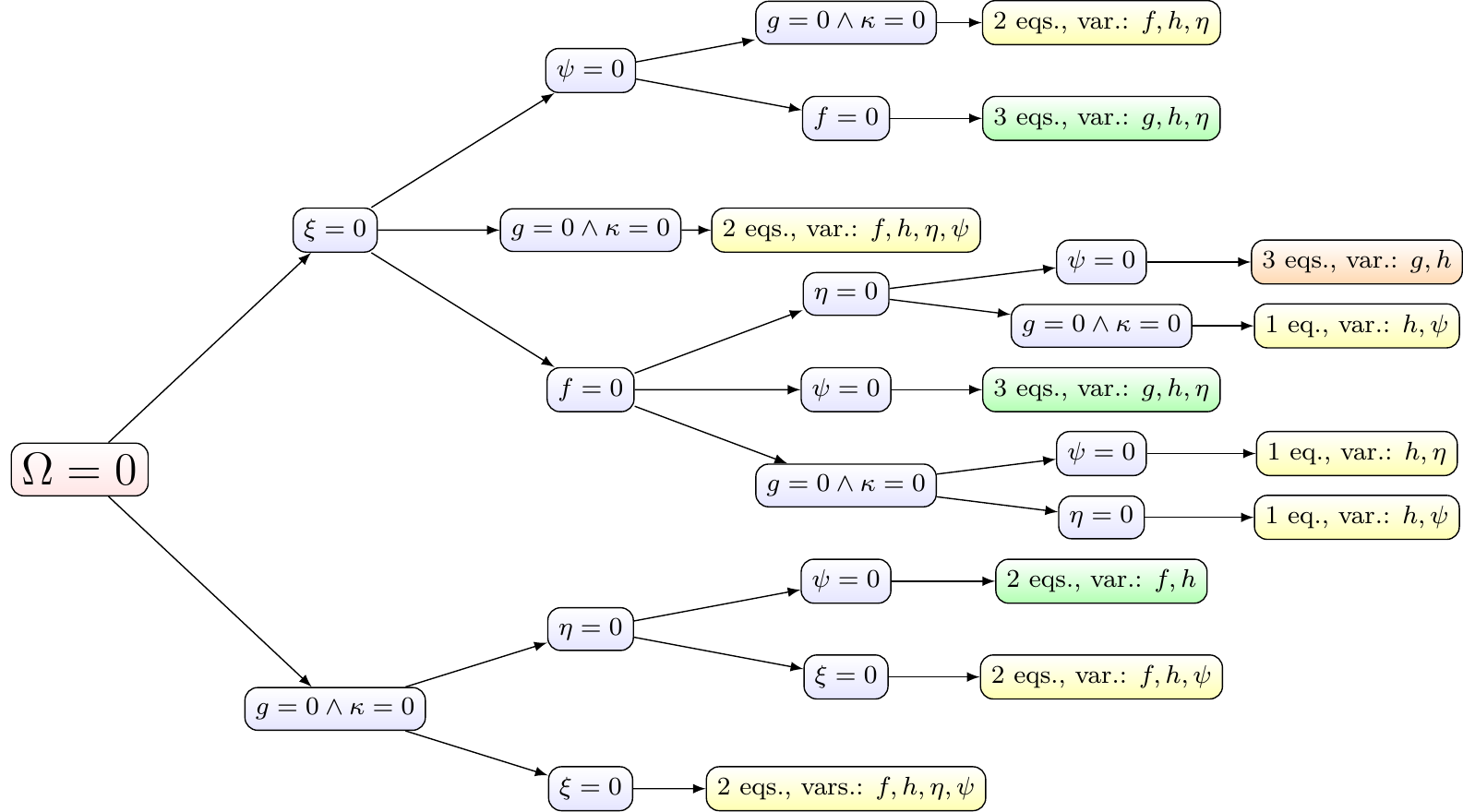}
\end{center}

The solutions obtained after exhausting the branches of the decision
trees are presented below, but they are grouped according to the
fields that are nontrivial. 

\subsection{Torsion-free limit}
\label{sec:org706c837}

Consider first the field equation \eqref{eq:offshell_feqs_torsion_free}.
Although at first sight the solution seems that spacetime is flat,
when working the expression carefully, one notice that the
requirements are that,
\begin{equation}
  \begin{aligned}
    f g & = 0,
    &
    \dot{g} - 2 g h - \kappa & = 0,
    &
    \dot{f} + 2 f h & = 0.
  \end{aligned}
  \label{eq:b8_wr_gamma}
\end{equation}

However, the extra equation obtained when we take the \emph{on-shell}
torsion-free limit, i.e. Ricci-flatness, is satisfied simultaneously
if and only if the connection is flat.

\subsubsection{Projectively-flat solutions}
\label{sec:org57d69b9}

A projectively-flat connection requires that
\begin{equation}
  \begin{aligned}
    gf & = 0
    &
    \dot{f} + 2fh & = 0
    &
    \dot{g} -2gh - \kappa & = 0
  \end{aligned}
\end{equation}

There are two \emph{branches} of solutions, with either \(g = 0\) or \(f =
0\). None of these \emph{branches} requires the connection to be flat.
Additionally, since \(h\) is a non-dynamical function, it is possible
to solve the field equations in terms of an \(h\)-parameter function.

The branch \(g = 0\), requires vanishing \(\kappa\). In this scenario,
the \(f\)-function is solved by
\begin{equation*}
  f(t) = C_f e^{-2 H(t)},
\end{equation*}
where \(H(t) = \int_{t_0}^{t} \de{\tau} h(\tau)\).

In the branch \(f = 0\), again the \(h\)-function plays the role of a
parameter function, and the \emph{dynamical} function is given by
\begin{equation*}
  g(t) = e^{2 H(t)} \left( C_g + \kappa \int_{t_0}^{t} \de{\tau} e^{-2 H(\tau)} \right),
\end{equation*}
with \(C_g\) the integration constant.

Note that the case \(f = g = 0\) leaves  the function \(h\)
undetermined, but it is compatible with both of the previous branches. 

\subsubsection{Flat solutions}
\label{sec:flat_solutions}
As mentioned previously, flat connections solve the field equations
obtained from the on-shell torsion-less limit of the polynomial affine
model of gravity.

From Eqs. \eqref{eq:homotropic_curvature}, the system of ordinary
differential equations defining isotropic and homogeneous affinely
connected flat spaces are
\begin{align*}
  f g & = 0,
  &
    \dot{g} - g h & = 0,
  &
    g h + \kappa & = 0,
  &
    \dot{h} + h^2 - f^2 & = 0,
  &  
    \dot{f} + 2 f h & = 0.
\end{align*}

As in the previous case, there are two branches of solutions. The
branch with \(f = 0\), the field equations are solved by 
\begin{equation}
  h(t) = \frac{1}{t + C_h},
  \quad
  g(t) = - \kappa ( t + C_h ),
\end{equation}
with \(C_h\) the integration constant of the equation for the
\(h\)-function.

The second branch requires \(g = 0 \land \kappa = 0\). Since the case \(f =
0\) is included the other branch, we restrict to \(f \neq 0\).
Therefore, the \(f\)-function can be integrated in terms of \(h\),
\begin{equation*}
  f(t) = e^{- 2 H(t)}.
\end{equation*}
Then, the field equation for \(h\) turns into a second order
differential equation. The solution of the system is given by

\begin{equation}
  \begin{aligned}
    h(t) & = \frac{t + c_1}{(t + c_1)^2 + 1},
    &
    f(t) & = \frac{1}{(t + c_1)^2 + 1}.
  \end{aligned}
  \label{eq:flat_conn_vanish_g}
\end{equation}

\subsubsection{Ricci-flat solutions}
\label{sec:orgaec0f4f}

Ricci-flat connections solve the field equations from the on-shell
torsion-free limit of polynomial affine gravity solely if the coupling
constant \(B_8\) is zero.

From Eqs. \eqref{eq:homotropic_ricci}, the system of ordinary
differential equations defining isotropic and homogeneous affinely
connected Ricci-flat spaces are
\begin{align}
  \label{eq:homotropic_riccati}
  \dot{h} + h^2 - f^2 & = 0,
  &
  \dot{g} + \kappa & = 0.
\end{align}

The solution for the differential equation for \(g\) is
\begin{equation*}
  g(t) = - \kappa t + C_g.
\end{equation*}

The first equation in Eq. \eqref{eq:homotropic_riccati} is a Riccati ordinary
differential equation, in which the \(f\)-function plays the role of
\emph{parameter} function. A well-known strategy to solve the Riccati
equation is to transform it into a second order linear differential
equation. The transformation \(u(t) = e^{H(t)}\), takes it onto
\begin{equation}
  \ddot{u} - f^2 u = 0.
  \label{eq:homotropic_riccati_u}
\end{equation}
Equation \eqref{eq:homotropic_riccati_u} can be immediately compared
with the one-dimensional time-independent Schrödinger equation, where
\(u\) would be the wave function, the \(f^2\)-function plays the role
of the quantum mechanical potential minus the energy eigenvalue. 

\subsection{Purely vectorial sector}
\label{sec:restriction_A}
The off-shell limit toward the purely vectorial sector is dominated by
the term whose coefficient is \(B_2\). The field equations are
identically satisfied, given that for \(\A_\mu = \eta(t)
\delta^0_\mu\) its field strength vanishes, i.e. the field equations
impose no restriction to the function \(\eta\).

On the other hand, the subsidiary condition coming from the on-shell
limit is
\begin{equation*}
  B_3 \, \dot{\eta} - B_1 \, \eta^2 = 0, 
\end{equation*}
whose solution is 
\begin{equation}
  \eta (t) = - \frac{B_3}{B_{1}} \frac{1}{t + C_\eta}.
\end{equation}

\subsection{Purely traceless torsion sector}
\label{sec:restriction_B}
In this sector, the field equations from the off-shell limit become
\(\psi^2 = \psi \xi = 0\), whose solution is driven by
\begin{equation}
  \psi = 0,
  \qquad
  \xi = \xi(t),
\end{equation}
whose solution is driven by \(\psi (t) = 0\), while \(\xi (t)\) remains as
an unknown function.

When one considers the on-shell limit, the system of field equations is
\begin{equation}
  \begin{aligned}
    \dot{\xi} & = 0,
    &
    \kappa & = 0,
    &
    \psi^2 & = 0,
    &
    -3 B_4 \, \psi \xi + B_6 \, \kappa & = 0,
  \end{aligned}
\end{equation}
which is solved by
\begin{align}
  \xi (t) & = C_\xi, & \psi(t) & = 0, & \kappa & = 0.
\end{align}

\subsection{Connection with vectorial torsion}
\label{sec:org93c0427}

The field equations, in the \emph{off-shell} limit, are those of
the non-Abelian Chern--Simons for the symmetric connection and the
Chern--Simons for the \(\A\)-field. They, however do not interact, and
additionally in the cosmological ansatz the field equation for \(\A\)
is automatically satisfied. Therefore, the cosmological models with
vectorial torsion in the off-shell limit, do not differ from those of
the torsion-free (off-shell) limit mixed---but non-interacting---with
an unconstrained vector field \(A_\mu = \eta(t) \delta^0_\mu\).

The field equations obtained in the \emph{on-shell} limit come with an
additional condition, which can be written as
\begin{equation*}
  B_6 \ri{\mu\nu}{}{} - B_3 \nabla_\mu \A_\nu + B_1 \A_\mu \A_\nu = 0.
\end{equation*}
The above expression yield two independent field equations,
\begin{align}
  \label{eq:feq4_vector_torsion_tt}
  - 2 B_6 \left( \dot{h} + h^2 - f^2 \right) - B_3 \dot{\eta} + B_1 \eta^2
  & = 0,
  \\
  \label{eq:feq4_vector_torsion_ii}
  B_6 \left( \dot{g} + \kappa \right) + B_3 \, g \eta
  & = 0.
\end{align}
Therefore the equations to solve are Eqs. \eqref{eq:b8_wr_gamma},
\eqref{eq:feq4_vector_torsion_tt} and \eqref{eq:feq4_vector_torsion_ii}.
From Eqs. \eqref{eq:b8_wr_gamma}, the branches structure is inherited,
i.e. \(g = 0 \lor f = 0\).

The branch of solutions with vanishing \(g\) requires that \(\kappa =
0\). The nontrivial field equations are then
Eq.~\eqref{eq:feq4_vector_torsion_tt} and 
\begin{equation*}
  \dot{f} + 2 f h = 0.
\end{equation*}
Since \(f(t) = \exp(- 2 H(t))\), Eq. \eqref{eq:feq4_vector_torsion_tt}
can be recasted as
\begin{equation*}
  - 2 B_6 \left( \dot{h} + h^2 - e^{-4 H(t)} \right) - B_3 \dot{\eta} + B_1 \eta^2 = 0.
\end{equation*}
From the last equation it is evident that a solution is given by the
flat connection \eqref{eq:flat_conn_vanish_g}, together with
\begin{equation*}
  \eta(t) = - \frac{B_3}{B_1} \frac{1}{t + C_\eta}.
\end{equation*}
Note in addition that for a generic function \(h\), unrelated to
\(\eta\), we can define 
\begin{equation*}
  \phi(t) = \frac{2 B_6}{B_1} \left( \dot{h} + h^2 - e^{-4 H(t)} \right), 
\end{equation*}
resulting in a Riccati equation for the \(\eta\)-function, which might
be transformed into the one-dimensional Schrödinger equation
\begin{equation*}
  \ddot{u} - \frac{B_3}{B_1} \phi(t) u(t) = 0,
\end{equation*}
with the change of variable, \(u(t) = \exp \left( - \tfrac{B_3}{B_1}
\Hi(t) \right)\) where \(\Hi(t) = \int_{t_0}^{t} \de{\tau}
\eta(\tau)\). Summarising, in this branch, given a \emph{parameter}
function \(h\), it is (in principle) possible to integrate the field
equations to determine the connection.

On the other hand, in the branch \(f = 0\), the nontrivial
differential equations to solve are
\begin{equation}
  \begin{aligned}
    B_1 \eta^2 - B_3 \dot{\eta} - 2 B_6 ( \dot{h} + h^2 ) & = 0,
    \\
    B_6 (\dot{g} + \kappa) + B_3 \eta g & = 0,
    \\
    \dot{g} - 2 g h - \kappa & = 0.
  \end{aligned}
  \label{eq:onshell_feqs_vect_tors_vanish_f}
\end{equation}
Solving for \(\eta\) and \(h\) from the last two expressions and
substituting into the first, one obtains a differential equation for
the \(g\)-function,
\begin{equation}
  \frac{B_6}{B_3} \left( B_3^2 - 2 B_1 B_6 \right) \frac{\dot{g} + \kappa}{g} = 0.
  \label{eq:g_feq_onshell_vect_tors}
\end{equation}

There are three branches of solutions:
\begin{enumerate}
\item For \(\dot{g} + \kappa = 0\), it follows that \(g = - \kappa t +
   C_g\), \(\eta = 0\) and \(h = \frac{\kappa}{2(\kappa t - C_g)}\).
Note that for \(\kappa = 0\) this solution is valid if \(C_g \neq
   0\), but the solution with \(g = 0\) requires that \(h =
   \frac{1}{t + C_h}\).
\item For \(B_6 = 0\) there are two kinds of solutions, both with
arbitrary \(h\) function: (i) \(g = 0 \land \kappa = 0\) and \(\eta
   = - \frac{B_3}{B_1} \frac{1}{t + C_\eta}\), or (ii) \(\eta = 0\)
and \(g(t) = e^{2 H(t)} \left( C_g + \kappa \int_{t_0}^{t}
   \de{\tau} e^{-2 H(\tau)} \right)\).
\item For \(B_3^2 = 2 B_1 B_6\), the solutions are parametrised by the
function \(g\), which is required to be nonvanishing and \(C^1\).
Hence, \(h = \frac{\dot{g} - \kappa}{2 g}\) and \(\eta = -
   \frac{B_3}{B_1} \frac{\dot{g} + \kappa}{g}\).
\end{enumerate}

\subsection{Connection with traceless torsion}
\label{sec:trl_torsion_solutions}
The field equations in the off-shell limit are obtained from
\eqref{eq:offshell_feq_gb1} and \eqref{eq:offshell_feq_gb2}, and yield 
\begin{align*}
  f g & = 0,
  &
    B_6 (2 g \psi + \dot{\xi}) - B_8 f g & = 0,
  &
    \dot{g} - 2 g h - \kappa & = 0,
  \\
    \dot{f} + 2 f h & = 0,
  &
    3 B_4 \psi^2 - 2 B_6 (\dot{h} + h^2 - f^2) & = 0,
  &
    - 3 B_4 \psi \xi + B_6 (\kappa + \dot{g}) & = 0.
\end{align*}
The \emph{extra} condition obtained from the on-shell limit is
\begin{equation*}
  \dot{\xi} + 2 g \psi = 0,
\end{equation*}
was already satisfied with the off-shell equations, i.e. there is no
difference between the off-shell and on-shell \(\A \to 0\) limit, for
the cosmological ansatz.

The solutions to the system of field equations are categorised in to
classes, those with \(g = 0\) or \(f = 0\).

The branch of solutions with \(g = 0\) requires that \(\kappa = 0\)
and either \(\psi = 0\) or \(\xi = 0\). On the one hand, for \(\psi =
0\), the field equations require that \(\xi = C_\xi\), while the
nontrivial equations are solved by \eqref{eq:flat_conn_vanish_g}, with
the difference that the torsion tensor field is nonvanishing. on the
other hand, for \(\xi = 0\) the field equations to be solved are
\begin{equation*}
  \dot{f} + 2 f h = 0 \quad
  \text{ and } \quad
  \dot{h} + h^2 - f^2 - \frac{3 B_4}{2 B_6} \psi^2 = 0.
\end{equation*}
Since there are three unknowns and only two equations, the solutions
are parametrised by one of the unknowns, e.g. \(\psi\). A simple
solution is obtained for \(\psi \propto f\), with results similar to
those in Eq. \eqref{eq:flat_conn_vanish_g}. Note that another solution
is given by \(f = 0\), in whose case the sole nontrivial field
equation is
\begin{dmath}
  \dot{h} + h^2 - \frac{3 B_4}{2 B_6} \psi^2 = 0.
\end{dmath}
This is a Riccati equation for \(h\), which is equivalent to a
one-dimensional Schrödinger equation, in which the function \(\psi\)
is the analogous to the quantum mechanical potential.

The system of equations for the branch with \(f = 0\) is
\begin{equation}
  \begin{aligned}
    \psi(t) + \frac{\dot{\xi}}{2 g} & = 0,
    &
    h (t) - \frac{\dot{g} - \kappa}{2 g} & = 0,
    \\
    - 3 B_4 \dot{\xi}^2 + 2 B_6 (\kappa^2 - \dot{g}^2 + 2 g \ddot{g}) & = 0,
    &
    2 B_6 (\kappa + \dot{g}) + \frac{3 B_4 \xi \dot{\xi}}{g} & = 0
  \end{aligned}
  \label{eq:eqs_gb_onshell_f0}
\end{equation}

The system of Eqs. \eqref{eq:eqs_gb_onshell_f0} is solved by
\begin{dgroup}
  \begin{dmath}
    \psi(t) = - \frac{\dot{\xi}}{2 g},
  \end{dmath}
  \begin{dmath}
    h (t) = \frac{\dot{g} - \kappa}{2 g},
  \end{dmath}
  \begin{dmath}
    \xi(t) = \sqrt{ C_\xi^2 - \frac{2 B_6}{3B_4} \bigg( g^2 + 2 \kappa \int g \de{t} \bigg) },
  \end{dmath}
  \begin{dmath}
    0 = g \ddot{g} \left( g^2 + 2 \kappa \int g \de{t} \right)
      + \kappa g^2 \left( \dot{g} + 1 \right)
      + \kappa \int g \de{t} \left( 1 - \dot{g}^2 \right)
      + \frac{3 B_4}{4 B_6} C_\xi \left( \dot{g}^2 - 2 g \ddot{g} - \kappa^2 \right).
  \end{dmath}
  % \label{eq:eqs_gb_onshell_f0}
\end{dgroup}
Particularly, for \(\kappa = C_\xi = 0\) the explicit solution is
given by\footnote{Formally, the field equations can be solved for \(C_\xi \neq
0\), but the solution for the \(g\)-function is expressed in terms of
the inverse of an hypergeometric function. Hence, we have omitted the
details of such solution.}
\begin{equation}
  \begin{aligned}
    f(t) & = 0,
    &
    g(t) & = C_m t + C_g,
    &
    h(t) & = \frac{C_m}{2 (C_m t + C_g)},
    \\
    \xi(t) & = \sqrt{ - \frac{2 B_6}{3 B_4}} \left( C_m t + C_g \right),
    &
    \psi(t) & = - \sqrt{ - \frac{B_6}{6 B_4}} \frac{C_m}{C_m t + C_g},
    &
    \eta(t) & = 0,
  \end{aligned}
  \label{eq:sol_GB_f0eta0_k0}
\end{equation}
where \(\sgn \left( B_4 \right) = - \sgn \left( B_6 \right)\).

For \(\kappa \neq 0\) a simplification of the field equations is to
propose \(g\) as a linear function of \(t\), yielding
\begin{equation}
  \begin{aligned}
    f(t) & = 0,
    &
    g(t) & = - \kappa t + C_g,
    &
    h(t) & = \frac{1}{t - C_g \kappa},
    \\
    \xi(t) & = C_\xi
    &
    \psi(t) & = 0,
    &
    \eta(t) & = 0.
  \end{aligned}
\end{equation}

\subsection{Restriction to torsional sector}
\label{sec:restriction_AB}
The solutions to the off-shell restriction to the torsional sector
require that either \(\psi = 0\) or \(\xi = 0\).\footnote{Note that \(\psi\)  and \(\xi\) cannot vanish at the same
time, since it would imply that \(\bt{}{}{} = 0\).} For \(\xi =
0\) the nontrivial field equation is the Riccati-like equation,
\begin{equation}
  B_1 \eta^2 - B_3 \dot{\eta} + 3 B_4 \psi^2 = 0,
\end{equation}
which (as mentioned before) is equivalent to a one-dimensional
Schrödinger equation whose potential is related to the function
\(\psi^2\). For \(\psi = 0\), the explicit solution is
\begin{equation}
  \psi(t) = 0,
  \quad
  \eta(t) = - \frac{B_3}{B_1} \frac{1}{t + C_\eta},
  \quad
  \xi(t) = C_\xi \left( t + C_\eta \right)^2.
\end{equation}

In the on-shell limit, the subsidiary conditions are \(\xi \eta = 0\),
\(\dot{\xi} = 0\) and \(\psi \eta = 0\). Hence, there is no solution
because the field equations require that \(\bt{}{}{} = 0\).

\subsection{Exceptional solutions to the whole model}
\label{sec:orgd57adfe}

The cosmological solutions to polynomial affine model of gravity in
three dimensions with all the fields turned on are \emph{exceptional},
since we are solving the seven dimensional system of field equations
\eqref{eq:cosmological_feqs}, with only six unknowns. Generically, the
consistency conditions of the system require that either some of the
functions vanish (worsening the well-being of the system) or a
relation between the parameters of the model.

A first example of these is given by the case \(g = \xi = \kappa =
0\), with nontrivial field equations
\begin{dgroup}
  \begin{dmath}
    \dot{f} + 2 f h = - \frac{B_3}{B_8} \eta \psi,
    \label{eq:f_param_eta_psi}
  \end{dmath}
  \begin{dmath}
    \dot{h} + h^2 - f^2 = \frac{1}{2 B_6} \left( B_1 \eta^2 - B_3
      \dot{\eta} + 3 B_4 \psi^2 \right).
    \label{eq:h_param_eta_psi}
  \end{dmath}
  \label{eq:fh_param_eta_psi}
\end{dgroup}
These field equations are solvable when all functions are inversely
proportional to \(t\).

Another branch of solutions is found when the coupling constants are
not all independent. As an example, for \(B_3^2 + 4 B_1 B_6 = 0\) the
field equations allow to decouple the functions \(g\) and
\(\xi\)---which are unconstrained---, from the equations for \(f\) and
\(\psi\), which should satisfy the differential equations
\eqref{eq:fh_param_eta_psi}. We were able to find two types of
solutions of this system of equations (which do not fall into the
previously presented categories).

\begin{itemize}
\item \textbf{Solution with \(B_3 = 0\):} With this condition, the solution is
characterised by the functions
\begin{equation}
  \begin{aligned}
    f(t) & = 0,
    &
    g(t) & = - \kappa t + C_g,
    &
    h(t) & = \frac{1}{t - \kappa C_g},
    \\
    \xi(t) & = \sqrt{- \frac{2 B_6}{3 B_4}} C_g,
    &
    \psi(t) & = 0,
    &
    \eta(t) & = \text{arbitrary},
  \end{aligned}
  \label{eq:exceptional_sol1_kn0}
\end{equation}
for \(\kappa \neq 0\). While for \(\kappa = 0\), the functions
defining the connection are
\begin{equation}
  \begin{aligned}
    f(t) & = 0,
    &
    g(t) & = C_m t + C_g,
    &
    h(t) & = \frac{C_m}{2 (C_m t + C_g)},
    \\
    \xi(t) & = \sqrt{- \frac{2 B_6}{3 B_4}} (C_m t + C_g),
    &
    \psi(t) & = \sqrt{- \frac{B_6}{6 B_4}} \frac{C_m}{C_m t + C_g},
    &
    \eta(t) & = \text{arbitrary}.
  \end{aligned}
  \label{eq:exceptional_sol1_k0}
\end{equation}
\item \textbf{Solutions for \(\kappa = 0\):} The system of field equations can be
solved for the ansatz \(g(t) = t^n\) with \(n \in \R - \set{-2,
  \left[\frac{1-\sqrt{33}}{4}, \frac{1+\sqrt{33}}{4}\right]}\), with
the addition condition
\begin{equation*}
  B_4 = - \frac{8 B_6^3 (n+2)}{3 B_8^2 \left(2 n^3-3 n^2-3 n+4\right)}.
\end{equation*}
The functions defining the connection are
\begin{equation}
  \label{eq:exceptional_sol2_k0}
  \begin{aligned}
    f(t) & = \frac{\sqrt{2 n^2 - n - 4} \sgn(B_8)}{2 t},
    &
    g(t) & = t^n,
    &
    h(t) & = \frac{n}{2 t},
    \\
    \xi(t) & = \frac{\sqrt{2 n^2 - n - 4} \sgn(B_8) t^n}{2 B_6},
    &
    \psi(t) & = \frac{(n - 1) \sqrt{2 n^2 - n - 4} \sgn(B_8)}{4 t B_6},
    &
    \eta(t) & = \frac{2 B_6}{B_3 t}.
  \end{aligned}
\end{equation}
\end{itemize}

\section{Discussion and conclusions}
\label{sec:conclusions}
In an attempt to bring gravity to the same footing than gauge
theories, we have proposed a model of gravity whose sole fundamental
field is the affine connection. Such model has been named \emph{polynomial
affine gravity}. Our model might be understood as a Schwarz
topological theory, in the sense that the metric plays no role in the
model building, similar to the case of Chern--Simons theories. 

The polynomial affine model of gravity has been built in three and
four dimensions (i.e. there are no \emph{ab initio} restrictions on the
dimension of the space, unlike for Chern--Simons theories), and
possesses attractive features. Firstly, the model is appealing for a
quantum theory of gravity, since all the terms in the action are
power-counting renormalisable, and in addition the lack of additional
invariant forms would forbid the existence of counter-terms. Secondly,
the absence of an energy scale, reflected by the fact that all the
coupling constants are dimensionless, appears as a hint (of a sort) of 
conformal invariance (at least at tree level).

Customarily, the conformal transformation is understood (in metric
gravitational models) as a point-wise scaling of the metric tensor
field, and the invariant curvature under these transformations is the
conformal Weyl tensor field, i.e. the \(g\)-traceless part of the
Riemann--Christoffel curvature. This notion, can be generalised
without evoking a metric. The idea is that \emph{self-parallel curves}
can be preserved under ``generalised'' transformations. These
are the \emph{projective} transformations, and the invariant curvature
under these transformations is the projective Weyl tensor
field.

In this article we focus in the three-dimensional version of the
polynomial affine model of gravity, where the action and therefore the
field equations are simpler than their four-dimensional analogous,
expecting the physical interpretation to be clearer. Note that in
comparison the three-dimensional action [see Eq. \eqref{eq:action}] is
determined by eight terms (including the Chern--Simons terms) while
the four-dimensional one is composed by twenty terms (disregarding
topological terms), unrelated through boundary terms.

Interestingly, the term of the action with coefficient \(B_3\) can be
re-written (up to boundary term) as, \(\nabla_\mu \A_\alpha
\bt{\beta}{\mu}{\gamma} \de{V}^{\alpha\beta\gamma}\). In this case,
the \(\bt{}{}{}\)-field would be non-dynamical, and therefore
interpreted as an auxiliary field. Furthermore, for the case with
\(B_4 = 0\), the \(\bt{}{}{}\)-field would be a Lagrangian multiplier.

It is worth noticing that unlike the three-dimensional version of
General Relativity, in an affine model the (projective) Weyl tensor
does not vanish necessarily, and thus there is room to novel
phenomenological effects. In particular, the field equation
\eqref{eq:feq_G} contains terms that can be related to the projective
Weyl curvature, and therefore its space of solutions might differ from
the one expected in General Relativity, even in the cases where the
field equations are alike, e.g. for flat or Ricci-flat manifolds.

The field equations derived from the action in Eq. \eqref{eq:action},
include a generalisation of the Einstein field equations,
\begin{equation}
  \ri{\mu(\nu}{\mu}{\rho)}
  = 
  - \frac{B_1}{B_6} \A_\nu \A_\rho
  + \frac{B_3}{B_6} \nabla_{(\nu} \A_{\rho)}
  - \frac{3 B_4}{2 B_6} \bt{\nu}{\mu}{\sigma} \bt{\rho}{\sigma}{\mu}
  = \tilde{T}_{\nu\rho},
  \label{eq:einstein_like}
\end{equation}
obtained by varying with respect to the \(\bt{}{}{}\)-field is the
\emph{analogous} to the Einstein equations written in the Ricci form.
Noticeable, the fact that---even for torsion-free truncation---the
field equations for the symmetric connection appear from the variation
with respect to other field, has been interpreted as a sign of the
non-uniqueness of the Lagrangian description of the system
\cite{hojman_privat}.

The Eq. \eqref{eq:einstein_like} represents a non-Riemannian
generalisation of the Einstein equations in the Ricci form, where the
right-hand side geometrically encodes what in General Relativity is
attributed to the presence of matter. However, \(\tilde{T}\) does not
admit a separation between material and geometrical contributions,
unlike its analogous form in General Relativity, which is expressed in
terms  of the energy-momentum tensor \(T_{\mu\nu}\) and its trace
\(T\), as 
\begin{equation*}
  \tilde{T}_{\nu\rho}^{GR} \propto T_{\nu\rho} - T g_{\nu\rho}.
\end{equation*}

Furthermore, when \(\tilde{T}\) is non-degenerated, it equips the
affine manifold with a torsion-descendent notion of metric, and hence
Eq. \eqref{eq:einstein_like} provides a notion of \emph{affine} Einstein
manifold. A nice feature of this torsion-descendent metric is that,
unlike the \emph{emergent metric} from Ref.
\cite{castillo-felisola20_emerg_metric_geodes_analy_cosmol}, it might be
well-defined even when the space is Ricci-flat.

However, since the affine connection is a less intuitive geometrical
object (in comparison with the metric), we analysed the possible
truncations of the model, i.e. sectors where only a subset of the
irreducible components of the connections are nontrivial.

Turning to the solutions of the field equations, we found the ansätze
of the three-dimensional affine connection compatible with the
cosmological principle. Firstly, we found that the symmetric
connection is determined by three functions,\footnote{When solving the differential equations obtained from the Lie
derivative of the connection, there is a fourth parametric function
(\(j\)) characterising the affine connection, but we show in
\ref{sec:t_reparam} that this parameter can be eliminated by a
reparametrisation on the \emph{time} coordinate.} \(f\), \(g\) and
\(h\). The \(f\) function has no analogous in a cosmological
Levi-Civita connection, and therefore it is a non-Riemannian
parameter. In addition, the \(g\) and \(h\) functions describe a
non-Riemannian cosmological geometry unless their could be
parametrised in terms of a scale factor, \(a = a(t)\), as
\begin{equation}
  g = a \dot{a}
  \quad
  \text{and}
  \quad
  h = \frac{\dot{a}}{a}.
  \label{eq:gh_param}
\end{equation}
Hence, in the torsion-free sector of the polynomial affine model of
gravity the non-Riemannian structure percolates the Einstein-like
equations if \(f \neq 0\) and/or \(g\) and \(h\) are not parametrised
as in Eq. \eqref{eq:gh_param}.

Before discussing the cosmological structure of the torsional fields,
we would like to briefly mention the geometrical meaning of the \(f\)
function in the cosmological ansatz of the symmetric connection.
Consider the symmetric part of the affine connection defining the
polynomial affine model of gravity, \(\ct{\mu}{\lambda}{\nu}\), and a
generic metric, \(g_{\mu\nu}\). Let \(\ct.{\mu}{\lambda}{\nu}\) the
Levi-Civita connection associated to the metric \(g\). The difference
between the two connections, \(\ct{\mu}{\lambda}{\nu} -
\ct.{\mu}{\lambda}{\nu}\), is a tensor defining the Weyl's \emph{congruent
transferences}.\footnote{These transformations might be also called \emph{congruent
transplantation}, which is the translation of the original German
vocable (kongruente Verpflanzung).} Such tensor does no have an analogous in
Riemannian geometry, since it is related to the non-metricity. The
cosmological ansatz for the connection in three dimensions admits
certain components of the \emph{transference} tensor, determined by the
function \(f\). Interestingly, in the cosmological ansatz in four
dimensions requires vanishing \emph{transference} tensor.

The (cosmological) torsion field is characterised by three functions,
\(\eta\), \(\xi\) and \(\psi\), the first one defines the
\(\Ag\)-field and the remaining two define the \(\bt{}{}{}\)-field. In
comparison with the four-dimensional case, which is characterised by
just two functions, the torsion tensor field is less restricted.
Noticeable, the same is true about the characterisation of the
(cosmological) symmetric connection, since in the four-dimensional
scenario it is determined (after the time reparametrisation) by solely
two functions.\footnote{We would like to stress that in Refs.
\cite{castillo-felisola18_beyond_einstein,castillo-felisola18_cosmol,castillo-felisola20_emerg_metric_geodes_analy_cosmol}
we were not aware of the \emph{time} reparametrisation, and therefore the
additional function turns the manipulation of the field equations into
a more cumbersome process.}

Using the cosmological ansatz, the Eq. \eqref{eq:einstein_like} is
written as
\begin{equation}
  \dot{h} + h^2 = \Xi(\text{non-Riemannian terms in } \ct*{}{}{}),
  \quad
  \left(
    \text{in GR: }
    \dot{H} + H^2 \propto (\rho + 3 p)
    \text{ with }
    H = \frac{\dot{a}}{a}
  \right)
  \label{eq:friedmann_equation}
\end{equation}
where \(h\) is one of the functions defining the symmetric connection
and its analogous in General Relativity, i.e. \(H\), is the Hubble
parameter. Eq. \eqref{eq:friedmann_equation} is a generalisation of the
Friedmann equation, where other geometric fields uphold what in
General Relativity would be interpreted as matter effects.
Generically, the \(\Xi\) function depends on all the other (i.e. non
\(h\)) functions characterising the affine connection, but it is
peculiar that the case where only \(f\) and \(h\) are nonvanishing,
the Eq. \eqref{eq:friedmann_equation} becomes a Riccati ordinary
differential equation 
\begin{equation*}
  \dot{h} + h^2 - f^2 = 0,
\end{equation*}
which can be expressed through the transformation \(u(t) = e^{H(t)}\),
where \(H(x) = \int_{x_0}^{x} \de{y} h(y)\), onto a one-dimensional
``time-independent'' Schrödinger equation, \(\ddot{u} - f^2 u = 0\),
where \(u\) would be the wave function, the \(f^2\)-function plays the
role of the quantum mechanical potential (with the energy eigenvalue
subtracted). Note that particularly in the flat and Ricci-flat cases
(coming from the torsion-free truncation) the Riccati equation with
constant \(f\) is analogous to the (acceleration) Friedmann equation
describing a Universe filled with a perfect fluid in a dark energy
dominated era, which in General Relativity is
\begin{equation*}
  \dot{h} + h^{2} - \frac{8 \pi G}{3} \rho_{DE} = 0.
\end{equation*}

In Sec. \ref{sec:scan} have found cosmological solutions to the field
equation of the polynomial affine model of gravity, characterised by
the system in Eq. \eqref{eq:cosmological_feqs}. We proceeded
systematically, scanning all the possible kinds of solutions. We noted
that the solutions split into two categories, depending on whether the
function \(\Omega\) from Eq. \eqref{eq:oo_omega} vanishes or not.
However, we re-classified the solutions according to the type of
truncation they belong to. It is worth mentioning that even though in
the classification of solutions we do not consider explicitly the case
with vanishing \(B_8\), this branch of solutions lying in the sector
\(\Omega = 0\), yields no additional solutions to those in Sec.
\ref{sec:scan}, e.g. the solutions with \(f = 0\) in Sec.
\ref{sec:flat_solutions} and the solutions in Sec.
\ref{sec:trl_torsion_solutions}. Furthermore, in the sector with \(4 B_1
B_6 + B_3^2 = 0\) we found solutions to the whole system of field
equations, where none of the functions determining the components of
the connection vanish.

Now, even though we have found \emph{affine} cosmological solutions to the
polynomial affine model of gravity in three dimensions, \emph{real life}
applications require the existence of a metric. In purely affine
models, although the fundamental field is the connection, it is
possible to define various types of (derived) metrics. Let us mention
four of these derived metrics:
\begin{enumerate}
\item The symmetric part of the Ricci tensor field, when it is
non-degenerated, serves as a metric. This was noticed very early
in the development of differential geometry (see for example
section 5 of Ref. \cite{eisenhart27_non_rieman}), and used recently
in Ref. \cite{castillo-felisola20_emerg_metric_geodes_analy_cosmol}.
A notable disadvantage is that interesting cases, such as Minkowski
and Schwarzschild, cannot be described using this notion of metric.
\item The quasi-Hodge dual of the \(\bt{}{}{}\)-field, i.e. \(T^{\mu\nu}
   = \frac{1}{2} \bt{\alpha}{\mu}{\beta} \epsilon^{\nu\alpha\beta}\),
is a symmetric \(\binom{2}{0}\)-tensor density. When \(T\) is
non-degenerated, it can be used as an \emph{inverse metric density},
similar to that used by Eddington, Einstein, Schrödinger and others
to build affine models of General Relativity (see Ref.
\cite{tonnelat14_einst}). In Ref.
\cite{castillo-felisola15_polyn_model_purel_affin_gravit} this
analogue is used to intuitively relate the three-dimensional action
of polynomial affine gravity with General Relativity nonminimally
coupled to the \(\A\)-field (see Eq. (9) of the referred article).
\item The construction \(\tt{\mu}{\lambda}{\rho}
   \tt{\nu}{\rho}{\lambda}\), defined from the torsion, is a symmetric
\(\binom{0}{2}\)-tensor field that serves (when non-degenerated) as
a metric. This tensor was introduced by Poplawski in Ref.
\cite{poplawski14_affin_theor_gravit}, and it is related to the
symmetric part of Eq. \eqref{eq:def_S_ricci1} (since the
\(\mathcal{S}\)-tensor is proportional to the torsion).
\item The symmetric part of \(\stt{\sigma\mu}{\sigma}{\nu}\) in Eq.
\eqref{eq:def_S_ricci1} can also be interpreted as a metric, when it
is non-degenerated.
\end{enumerate}

Each of the examples above serves to endow the affine manifold with a
metric structure with respect to which we could measure geodesic
distances and compare with the parallel transport obtained using the
symmetric connection. 

Only a few of the solutions explicitly presented in the paper admit
non-degenerated metrics. From the solution to the coupled
system \(\Gamma\)--\(\A\), the case with vanishing \(f\) and \(B_6\) [see the
second type of solutions after Eq. \eqref{eq:g_feq_onshell_vect_tors}]
possesses a \emph{Ricci} metric as long as the arbitrary function \(h\) is
not the reciprocal of \(t\), while the case with vanishing \(f\) and
\(B^2_3 = 2 B_1 B_6\) [see the third type of solutions after Eq.
\eqref{eq:g_feq_onshell_vect_tors}] possesses a metric of the fourth
kind, as long as \(g \neq - \kappa t + C_g\). The solutions of the coupled system
\(\Gamma\)--\(\bt{}{}{}\) with \(\kappa = 0\) [see Eq.
\eqref{eq:sol_GB_f0eta0_k0}] and the exceptional cases with vanishing
\(\kappa\) [see Eqs. \eqref{eq:exceptional_sol1_k0} and
\eqref{eq:exceptional_sol2_k0}], possess the first three types of metric
described above, as long as \(C_m \neq 0\) and \(n \neq 0\). In
addition, one can endow the exceptional solutions with the fourth kind
of metric, with the constrain that \(\eta\) is not a constant function
in Eqs. \eqref{eq:exceptional_sol1_kn0} and \eqref{eq:exceptional_sol1_k0}.

The above opens the discussion of the background independence of the
gravitational model, which is explicitly broken in General Relativity
due to the presence of the metric in the Einstein--Hilbert action.

\appendix

\section{Notions of non-Riemannian geometry}
\label{app:notation}
The aim of this appendix is to provide a short summary of results in
non-Riemannian geometry, and also intends to fix the notation used
through the development of the paper. Readers interested in the
subjects are recommended to review the classical texts by L. Eisenhart
\cite{eisenhart27_non_rieman} and J. Schouten \cite{schouten13_ricci},
the final chapter of the book by Synge and Schild
\cite{synge78_tensor}, the book by P. Gilkey and collaborators
\cite{stana13_applic_affin_weyl_geomet}, and articles like Refs.
\cite{siqueira18_topic_non_rieman_geomet,iosifidis19_metric_affin_gravit_cosmol_torsion,klemm20_einst_manif_with_torsion_nonmet}. 

A \(d\)-dimensional affine manifold \((\Mi,\hat{\nabla})\) is a
\(d\)-dimensional differential manifold \(\Mi\) equipped with a linear
connection \(\hat{\nabla}\). The linear connection is determined by
their \(d^3\)-independent components \(\ct*{\mu}{\lambda}{\rho}\),
such that 
\begin{equation*}
  \hat{\nabla}_\mu = \partial_\mu + \ct*{\mu}{\bullet}{\bullet}.
\end{equation*}

Since an affine structure does not require the existence of a metric
tensor field, the affine connection admits a decomposition in their
lower indices, into their symmetric and skew-symmetric parts,
\begin{equation*}
  \ct*{\mu}{\lambda}{\rho} = \ct*{(\mu}{\lambda}{\rho)} + \ct*{[\mu}{\lambda}{\rho]} = \ct{\mu}{\lambda}{\rho} + \stt{\mu}{\lambda}{\rho}.
\end{equation*}
The symmetric component of the connection remains as a connection,
which we denote by simply \(\ct{\mu}{\lambda}{\rho}\), while the
skew-symmetric component is a tensor field proportional to the torsion
tensor field (explicitly, it is twice the tensor field).

The curvature of a connection, defined by
\begin{equation}
  \hat{\Ri}(X,Y) Z = \left( \hat{\nabla}_X \hat{\nabla}_Y - \hat{\nabla}_Y \hat{\nabla}_X - \hat{\nabla}_{[X,Y]} \right) Z,
  \label{eq:def_curvature_hat}
\end{equation}
with \(X\), \(Y\) and \(Z\) vector fields, can be written in
components as
\begin{equation}
  \ri*{\mu\nu}{\lambda}{\rho}
  = \partial_\mu \ct*{\nu}{\lambda}{\rho}
  - \partial_\nu \ct*{\mu}{\lambda}{\rho}
  + \ct*{\mu}{\lambda}{\sigma} \ct*{\nu}{\sigma}{\rho}
  - \ct*{\nu}{\lambda}{\sigma} \ct*{\mu}{\sigma}{\rho}.
  \label{eq:curvature_affine_connection}
\end{equation}
The curvature tensor field is skew-symmetric in the first couple of
indices, and therefore the contra-variant index can be contracted in
two independent ways, \(\ri*{\lambda\nu}{\lambda}{\rho}\) and
\(\ri*{\mu\nu}{\lambda}{\lambda}\), referred to as the first and
second Ricci curvatures. Customarily, the first Ricci curvature is
simply called Ricci tensor field, while the second Ricci curvature is
also referred to as homothetic curvature or \emph{trace of the curvature}.
Note that when restricting oneself to Riemannian connections the Ricci
tensor field is symmetric and the trace of the curvature vanishes, but
for generic affine connections these properties do not hold.

The curvature in Eq. \eqref{eq:curvature_affine_connection} can be
decomposed in terms of the symmetric connection
\(\ct{\mu}{\lambda}{\rho}\) and the tensor
\(\stt{\mu}{\lambda}{\rho}\), yielding
\begin{dmath}
  \ri*{\mu\nu}{\lambda}{\rho}
  =
  \ri{\mu\nu}{\lambda}{\rho}
  + \hat{\nabla}_\mu \stt{\nu}{\lambda}{\rho}
  - \hat{\nabla}_\nu \stt{\mu}{\lambda}{\rho}
  - \stt{\mu}{\lambda}{\sigma} \stt{\nu}{\sigma}{\rho}
  + \stt{\nu}{\lambda}{\sigma} \stt{\mu}{\sigma}{\rho}
  - 2 \stt{\mu}{\sigma}{\nu} \stt{\sigma}{\lambda}{\rho}
  =
  \ri{\mu\nu}{\lambda}{\rho}
  + \nabla_\mu \stt{\nu}{\lambda}{\rho}
  - \nabla_\nu \stt{\mu}{\lambda}{\rho}
  + \stt{\mu}{\lambda}{\sigma} \stt{\nu}{\sigma}{\rho}
  - \stt{\nu}{\lambda}{\sigma} \stt{\mu}{\sigma}{\rho}
  =
  \ri{\mu\nu}{\lambda}{\rho}
  + \stt{\mu\nu}{\lambda}{\rho}.
  \label{eq:split_curvature_affine_connection}
\end{dmath}

Given the split of the curvature of the affine connection in Eq.
\eqref{eq:split_curvature_affine_connection}, we can analyse the
contractions of their components. Let us take first the curvature of
the symmetric connection,
\begin{equation*}
  \ri{\mu\nu}{\lambda}{\rho}
  =
  \partial_\mu \ct{\nu}{\lambda}{\rho} - \partial_\nu \ct{\mu}{\lambda}{\rho}
  + \ct{\mu}{\lambda}{\sigma} \ct{\nu}{\sigma}{\rho} - \ct{\nu}{\lambda}{\sigma} \ct{\mu}{\sigma}{\rho}.
\end{equation*}
The Ricci tensor is given by
\begin{equation}
  \ri{\nu\rho}{}{}
  =
  \ri{\lambda\nu}{\lambda}{\rho}
  =
  \partial_\lambda \ct{\nu}{\lambda}{\rho} - \partial_\nu \ct{\rho}{\lambda}{\lambda}
  + \ct{\sigma}{\lambda}{\lambda} \ct{\nu}{\sigma}{\rho} - \ct{\nu}{\lambda}{\sigma} \ct{\rho}{\sigma}{\lambda},
  \label{eq:ricci_symmetric_connection}
\end{equation}
which splits into symmetric and skew-symmetric parts,
\begin{align*}
  \ri{(\nu\rho)}{}{}
  & =
  \partial_\lambda \ct{\nu}{\lambda}{\rho} - \frac{1}{2} \left( \partial_\nu \ct{\rho}{\lambda}{\lambda} + \partial_\rho \ct{\nu}{\lambda}{\lambda} \right)
  + \ct{\sigma}{\lambda}{\lambda} \ct{\nu}{\sigma}{\rho} - \ct{\nu}{\lambda}{\sigma} \ct{\rho}{\sigma}{\lambda},
  \\
  \ri{[\nu\rho]}{}{}
  & =
  - \frac{1}{2} \left( \partial_\nu \ct{\rho}{\lambda}{\lambda} - \partial_\rho \ct{\nu}{\lambda}{\lambda} \right).
\end{align*}
From the transformation of the connection under diffeomorphisms, one
finds that the trace of the symmetric connection, \(\ct{\mu}{}{} =
\ct{\mu}{\lambda}{\lambda}\), transforms like
\begin{equation*}
  \Gamma'_{\mu}
  =
  \ct{\alpha}{}{} \frac{\partial x^\alpha}{\partial x^{\prime\mu}}
  + \partial_\mu \ln \mathfrak{d},
\end{equation*}
with \(\mathfrak{d} = \det \frac{\partial x^\alpha}{\partial
x^{\prime\nu}}\) is the scalar density defined by the determinant of
the transformation. Therefore, the skew-symmetric part of the Ricci
tensor is given by the curl of a vector
\begin{equation}
  \ri{[\mu\nu]}{}{} = \partial_\mu a_\nu - \partial_\nu a_\mu,
  \quad
  \text{ with }
  a_\mu = a_\mu{}^\lambda{}_\lambda.
  \label{eq:skew_ricci_symmetric_conn}
\end{equation}
The tensor \(a_\mu{}^\lambda{}_\rho\) is defined by the difference
between the connection \(\ct{\mu}{\lambda}{\rho}\) and a symmetric
connection of reference. In order to illustrate this point, we show
two cases: (i) if one chooses the Levi-Civita connection as reference,
\(a_\mu{}^\lambda{}_\rho\) is the \emph{contorsion} tensor; and (ii) if the
reference is given by the parameters of the parallel displacement in
an Euclidean manifold, then \(a_\mu{}^\lambda{}_\rho = \frac{\partial
x^\lambda}{\partial y^{\prime\alpha}}  \frac{\partial
y^{\prime\alpha}}{\partial x^{(\mu} \partial x^{\rho)}}\).

The second contraction of the curvature, i.e. the trace of curvature,
is given by
\begin{equation*}
  \ri{\mu\nu}{\lambda}{\lambda}
  =
  \partial_\mu \ct{\nu}{}{} - \partial_\nu \ct{\mu}{}{}
  =
  - 2 \ri{[\mu\nu]}{}{}.
\end{equation*}
The last step might be obtained from the algebraic Bianchi identity
for a torsion free connection.

The second term coming from the splitting of the curvature tensor,
\begin{equation}
  \stt{\mu\nu}{\lambda}{\rho}
  = \nabla_\mu \stt{\nu}{\lambda}{\rho}
  - \nabla_\nu \stt{\mu}{\lambda}{\rho}
  + \stt{\mu}{\lambda}{\sigma} \stt{\nu}{\sigma}{\rho}
  - \stt{\nu}{\lambda}{\sigma} \stt{\mu}{\sigma}{\rho},
  \label{eq:def_S_curvature}
\end{equation}
admits two contractions
\begin{equation}
  \label{eq:def_S_ricci1}
  \stt{\mu\nu}{\mu}{\rho}
  = \nabla_\mu \stt{\nu}{\mu}{\rho}
  + \nabla_\nu \stt{\rho}{}{}
  - \stt{\sigma}{}{} \stt{\nu}{\sigma}{\rho}
  + \stt{\nu}{\mu}{\sigma} \stt{\rho}{\sigma}{\mu},
\end{equation}
and
\begin{equation}
  \label{eq:def_S_ricci2}
  \stt{\mu\nu}{\lambda}{\lambda}
  =
  \nabla_\mu \stt{\nu}{}{} - \nabla_\nu \stt{\mu}{}{}
  =
  \partial_\mu \stt{\nu}{}{} - \partial_\nu \stt{\mu}{}{}.
\end{equation}
In the above equations we have introduced the vector \(\stt{\mu}{}{} =
\stt{\mu}{\lambda}{\lambda} = - \stt{\lambda}{\lambda}{\mu}\).

Note that both terms of the homothetic curvature of the linear
connection, \(\ct*{}{}{}\), is the curl of \(\stt{\mu}{}{} - a_\mu\)
and thus invariant under the addition of a gradient, i.e.
\(\stt{\mu}{}{} - a_\mu \mapsto \stt{\mu}{}{} - a_\mu + \partial_\mu
\phi\). This reflects a \emph{gauge} redundancy in the trace of the
curvature, which is inherited from the freedom of choosing a
connection of reference to define the \(a\) tensor.

Let us turn to the Bianchi identities. The torsion of the affine
connection is defined as
\begin{equation}
  \hat{\tor}(X, Y) = \hat{\nabla}_X Y - \hat{\nabla}_Y X - [X,Y],
  \label{eq:def_torsion_hat}
\end{equation}
and its derivative is
\begin{equation}
  \hat{\nabla}_Z (\hat{\tor}(X, Y))
  = \hat{\nabla}_Z \hat{\tor}(X, Y)
  + \hat{\tor}(\hat{\nabla}_Z X, Y)
  + \hat{\tor}(X, \hat{\nabla}_Z Y).
  \label{eq:derivative_of_torsion}
\end{equation}
The derivative of the vectors in last two terms of Eq.
\eqref{eq:derivative_of_torsion}, are expressible in terms of the
torsion, since
\begin{equation}
  \hat{\tor}(\hat{\tor}(X,Y), Z)
  = \hat{\tor}(\hat{\nabla}_X Y, Z)
  + \hat{\tor}(Z, \hat{\nabla}_Y X)
  - \hat{\tor}([X,Y], Z).
  \label{eq:torsion_along_torsion}
\end{equation}
The algebraic Bianchi identity is obtained by adding the cyclic
permutation of the vectors \(X\), \(Y\) and \(Z\), which shall be
denoted by the operator \(\mathfrak{S}_{X,Y,Z}\). Therefore, from Eq.
\eqref{eq:torsion_along_torsion} one gets
\begin{equation}
  \mathfrak{S}_{X,Y,Z} \left( \hat{\Ri}(X, Y, Z)
  - \hat{\tor}(\hat{\tor}(X,Y), Z)
  - \hat{\nabla}_X \hat{\tor}(Y, Z) \right)
  = 0.
  \label{eq:algebraic_bianchi_identity}
\end{equation}

The differential Bianchi identity is obtained from the derivative of
the curvature in Eq. \eqref{eq:def_curvature_hat},
\begin{equation*}
  \hat{\nabla}_Z (\hat{\Ri}(X, Y) W)
  = \hat{\nabla}_Z \hat{\Ri}(X, Y) W
  + \hat{\Ri}(\hat{\nabla}_Z X, Y) W
  + \hat{\Ri}(X, \hat{\nabla}_Z Y) W
  + \hat{\Ri}(X, Y) \hat{\nabla}_Z W,
\end{equation*}
after expressing the derivatives of the vectors (others than
\(W\)) in terms of the torsion tensor, and the application of the
cyclic permutation operator,
\begin{equation}
  \mathfrak{S}_{X,Y,Z} \left( \hat{\nabla}_Z \hat{\Ri}(X, Y) W
  + \hat{\Ri}(\hat{\tor}(X, Y), Z) W \right)
  = 0.
  \label{eq:differential_bianchi_identity}
\end{equation}
In order to get the last expression one uses that
\begin{equation*}
  \comm{\hat{\nabla}_Z}{\hat{\Ri}(X, Y)} W
  - \hat{\Ri}([X,Y], Z)
  = [\hat{\nabla}_Z, [\hat{\nabla}_X, \hat{\nabla}_Y]] W
  + \hat{\nabla}_{[[X,Y], Z]} W,
\end{equation*}
and the action of the cyclic permutation operator on it vanishes due
to the Jacobi identity of the Lie bracket and the commutator.

\section{Reparametrisation of time coordinate}
\label{sec:t_reparam}
Under a change of coordinates, \(x^{\prime a} = x^{\prime a}(x^i)\), the
component of the connection transform as
\begin{equation}
  \label{eq:transformation_connection}
  \frac{\partial^2 x^i}{\partial x^{\prime a} \partial x^{\prime b}}
  +
  \ct{j}{i}{k} \frac{\partial x^j}{\partial x^{\prime a}} \frac{\partial x^k}{\partial x^{\prime b}}
   = 
   \Gamma^{\prime \, c}_{a \; b} \frac{\partial x^i}{\partial x^{\prime c}}.
\end{equation}

Given the generic form of the isotropic and homogeneous connection,
Eq. \eqref{eq:homotropic_connection}, one notices that the function
\(j\) has no dynamics in the curvature tensors. A natural question is
whether there is a reparametrisation of the time coordinate that
allows to set \(j = 0\). Since this function comes from the component
\(\ct{0}{0}{0}\), consider a transformation of the form,
\begin{align*}
  t' & = t'(t), & r' & = r & \varphi' & = \varphi.
\end{align*}

Equation \eqref{eq:transformation_connection} for \(a = b = c = 0\), and
\(\Gamma^{\prime \, 0}_{0 \; 0} = 0\), yields
\begin{equation*}
  \frac{\partial^2 t}{\partial t^{\prime 2}} + j \cdot \left( \frac{\partial t}{\partial t'} \right)^2 = 0,
\end{equation*}
which can be written as a total derivative,
\begin{equation*}
  \frac{1}{X} \partial_{t'} \left( X \partial_{t'} t \right) = 0,
\end{equation*}
for \(j = \frac{1}{X} \partial_t X\), or equivalently \(X = e^{\int
\de{t} j(t)} = e^{J(t)}\). From here, one have that
\begin{equation*}
  t' = \int \de{t} e^{J(t)}.
\end{equation*}

With the above transformation, it can be checked with ease that the
effect of the time reparametrisation is a scaling of the other three
functions of time entering in the connection, and thus \(f(t) \mapsto
f(t')\), \(g(t) \mapsto g(t')\) and \(h(t) \mapsto h(t')\).

\begin{acknowledgements}
  The work OCF and ARZ is sponsored by the ``Centro Cient\'ifico y
  Tecnol\'ogico de Valpara\'iso'' \mbox{(CCTVal)}, funded by the
  Chilean Government through the Centers of Excellence Base Financing
  Program of Agencia Nacional de Investigaci\'on y Desarrollo (ANID),
  by grant ANID PIA/APOYO AFB180002. This work was funded by ANID
  Millennium Science Initiative Program \verb+ICN2019_044+, and
  benefited from the grant \verb+PI_LI_19_02+ from the Universidad
  T\'ecnica Federico Santa Mar\'ia. The work of JP is funded by
  USM-PIIC grant number \verb+059/2018+.
\end{acknowledgements}

\bibliographystyle{spphys}       % APS-like style for physics
\bibliography{References}   % name your BibTeX data base

%% % Non-BibTeX users please use
%% \begin{thebibliography}{}
%% %
%% % and use \bibitem to create references. Consult the Instructions
%% % for authors for reference list style.
%% %
%% \bibitem{RefJ}
%% % Format for Journal Reference
%% Author, Article title, Journal, Volume, page numbers (year)
%% % Format for books
%% \bibitem{RefB}
%% Author, Book title, page numbers. Publisher, place (year)
%% % etc
%% \end{thebibliography}

\end{document}